\documentclass[structabstract]{aa}
\usepackage{graphicx}
\usepackage{amsmath}
\usepackage{amsfonts}
\usepackage{amssymb}
\usepackage{natbib}
\newcommand{\varv}{v}

\bibpunct{(}{)}{;}{a}{}{,}
\begin{document}
   \title{Diagnostics of electron beam properties from the simultaneous hard X-ray and microwave emission in the 10 March 2001 flare}

     \author{V.V. Zharkova
          \inst{1},
          N.S. Meshalkina\inst{2}, L.K. Kashapova\inst{2}, A.A. Kuznetsov\inst{3}$^{,}$\inst{2}, A.T. Altyntsev\inst{2}
                  }


\institute{Department of Mathematics, University of Bradford, Bradford BD7 1DP, UK\\
    \email{v.v.zharkova@brad.ac.uk} \and
    Institute of Solar-Terrestrial Physics,
           Siberian Branch of the Russian Academy of
           Sciences,
           P.O. Box 4026, Irkutsk 33, 664033, Russia\\
           \email{nata@iszf.irk.ru}\and
             Armagh Observatory, Armagh BT61 9DG, Northern Ireland\\
             \email{aku@arm.ac.uk}}

   \date{Received September 15, 1996; accepted March 16, 1997}

  \abstract
{ Microwave (MW) and hard X-ray (HXR) data are thought to be
powerful means for investigating the mechanisms of particle
acceleration and precipitation in solar flares reflecting
different aspects of electrons interaction the ambient particles in a presence of
magnetic field. Simultaneous simulation of HXR and
MW emission with the same populations of electrons is still a
great challenge for interpretation of observations in real events.
Recent progress in simulations of particle kinetics with
time-dependent Fokker--Planck (FP) approach offers an
opportunity to produce such the interpretation.}
{ In this paper we apply the FP kinetic model of
precipitation of electron beam with energy range from 12 keV to
1.2 MeV to the interpretation of X-ray and microwave emissions
observed in the flare of 10 March 2001.  }
{ The theoretical HXR and MW emissions were calculated by
using the distribution functions of electron beams found by
solving time-dependent Fokker--Planck approach in a converging
magnetic field (Zharkova at al., 2010; Kuznetsov and
Zharkova, 2010) for anisotropic scattering of beam electrons
on the ambient particles in Coloumb collisions and Ohmic losses.  }
{ The simultaneous observed HXR photon spectra and frequency
distribution of MW emission and polarization were fit by those
simulated from FP models which include the effects of electric
field induced by beam electrons and precipitation into a
converging magnetic loop. Magnetic field strengths in the
footpoints on the photosphere were updated with newly calibrated
SOHO/MDI data. The observed HXR energy spectrum above 10 keV is
shown to be a double power law which was fit precisely by the
photon HXR spectrum simulated for the model including the
self-induced electric field but without magnetic convergence. The
MW emission simulated for different models of electron
precipitation revealed a better fit(above $90\%$ confidence level)
to the observed distribution at higher frequencies for the models
combining collisions and electric field effects with a moderate
magnetic field convergence of 2. The MW simulations were able to
reproduce closely the main features of the MW emission observed at
higher frequencies: the spectral index, the frequency of peak
intensity and the frequency of the MW polarisation reversal while
at lower frequencies the simulated MW intensities are lower than
the observed ones.  }
   {}

\authorrunning{Zharkova et al.}
\titlerunning{Diagnostics of the beam anisotropy}
  \maketitle
%
%


\section{Introduction}
Spatial configurations of flaring sites in hard X-ray (HXR) and
microwave (MW) emissions are highly variable, they may be
formed by a single loop with one coronal source and two footpoints
\citep{mas94,kun01a} or several sets of loops with a few
footpoints and coronal sources \citep{hanaoka96,kun01b} leading to
different models for different types of flaring events (see the
review by \citet{mel99}).

Simultaneous observations of hard X-ray emission and microwaves in
footpoints of solar flares often show their close temporal
correlation pointing out to their common origin \citep{asc05,
bas98}. There is a high likelihood that these emissions are
produced by the same population of non-thermal electrons
\citep{kun85, kun01b, kun01a, kun04, vil02, wil03, kun09}.

However, there are substantial discrepancies in the locations of
these HXR and MW sources observed in the same flare
--- e.g., they are often separated from each other, and
the areas covered by each emission are substantially different.
Observations often suggest that the sources of MW and HXR
emission may be separated in depths within a flaring loop
\citep{kun85,tak95,sui02}. The intensities of HXR and MW
emissions in the opposite legs of the same flaring loop
can be strongly anti-correlated being in one footpoint of the same
loop higher in MW and lower in HXR emission, while in the
other footpoint being higher in HXR and lower in MW emission
\citep{kun85,tak95,kun04,gre08}. In addition, there is sometimes a
delay between HXR and MW emissions occurring in
the same flare \citep{kun95,kun04,sui02}.

MW emission demonstrates a high variability of polarization from
a few percent \citep{alt00,fle03a, fle03b} up to 50--100\%
\citep{dul85,alt00,lee00}. Unlike HXR emission, which properties
are controlled by the parameters of emitting electrons and ambient plasma particles  and  Compton backscattering, i.e. albedo effects \citep{kon06} , MW
emission is determined in addition by various radiative
transfer effects. Nevertheless,  general perception of MW polarization
is that highly collimated  electron beams tend to produce MW
emission with a higher polarization degree \citep{fle03a,fle03b,Melnikov08,fle2010}.

\begin{figure*}
        \centering
   \includegraphics[width=17cm]{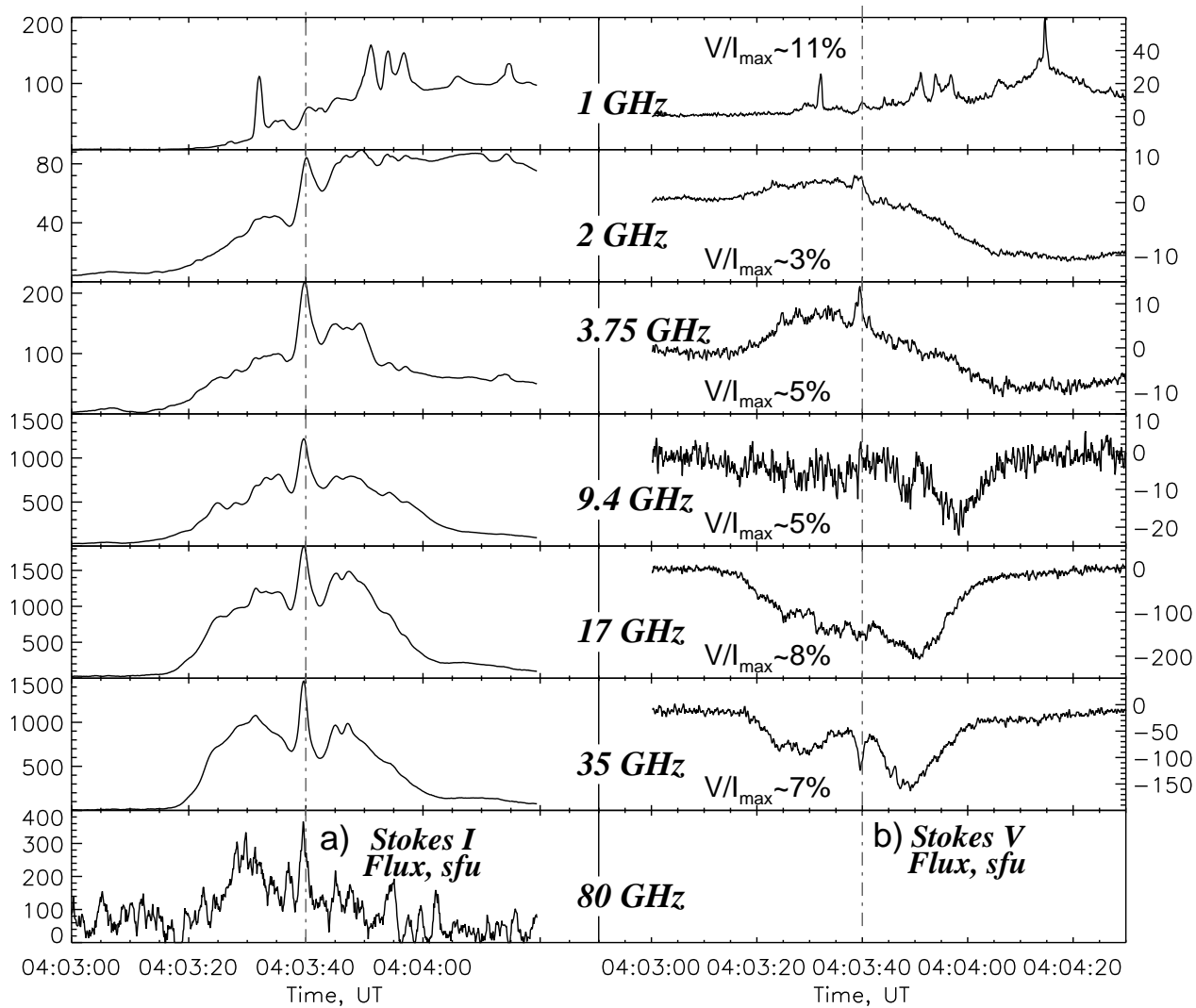}
\caption{The 10 March 2001 event. Time profiles of the average
brightness temperature at 1--80 GHz (NoRP) in total intensity (a)
and polarization (b). (V/I)$_{\max}$ is the maximum degree of
polarization at the central peak (04:03:40). The numbers in Y axes
are in the solar flux units (sfu).}
   \label{f-norp}
 \end{figure*}

Moreover, the resulting photon spectra of HXR and MW emission
produced during flares are also significantly different. In
particular, HXR emission from strongest flares often has the
elbow-type spectra with double power-law energy distributions
\citep{lin03,hol03} and noticeable differences between the
spectral indices at lower (below 70 keV) and higher energies. MW
distribution in frequency \citep{dul85,bas98} has a maximum with a
gradual decrease of the intensity towards lower frequencies
(optically thick emission) and a negative power law distribution
towards higher frequencies (optically thin emission). The peak
frequency usually varies from 3 to 20 GHz for different flares or
different stages of the same flare \citep{Nita04, Melnikov08}.

These differences clearly indicate different transport scenarios
for high-energy particles precipitating into the footpoints of a
flaring loop. The mechanisms of transport affecting HXR and MW
emissions are well known to be substantially different: MW
radiation is related to gyrosynchrotron emission of high-energy
electrons with energies from few tens keV \citep{kun01b} up to
several MeV \citep{bas99,kun04}, while HXR radiation is often
produced by the electrons with much lower energies from 10 to 300
keV \citep[see, for example,][]{lin03,hol03}. In spite of a large
number of simulations for the interpretation of HXR and
MW emission with thick target models \citep[see, for example, reviews][and references therein]{fle03a,fle03b,fle2010,kru08,krucker2010}, so far there are only a few simulations \citep{Melnikov08,kuz10} considering the problem of particle transport in other than collisional energy loss mechanisms and their effects on HXR and MW emission.

However, recent kinetic models calculated electron
distribution functions numerically by solving the time-dependent
Fokker--Planck kinetic equation for precipitation of electron
beams into a flaring atmosphere with converging magnetic field.
Beam electrons with an energy range from 12 keV to 10 MeV are
assumed to lose their energy and change propagation directions in
collisions and due to Ohmic losses in the electric field induced
by the beam \citep{dia88,zha95,zha06,bat2008} and due to scattering in converging magnetic field \citep{lea81,mcc92a,mcc92b,siversky09}.
These solutions were applied for the explanation of HXR
emission produced by solar flares which provided a very good fit
for HXR intensities \citep{mcc92b}, directivity and polarization \citep{zha10}.
Recent extension of these solutions to the interpretations of MW emission from solar flares
appears to produce promising results in closer fitting to observations \citep{kuz10}.

These kinetic models can be further tested by the simultaneous
interpretation of HXR and MW emission observed in the same flare.
For this purpose we selected the flare of the 10 March 2001 that
has been extensively studied \citep{Liu2001, ding2003, uddin2004,
chandra2006, altyntsev2008,Melnikov08} because of the following reasons: 1)
it had produced both kinds, HXR and MW emissions, MW with a
pronounced polarization; 2) it was a rather powerful flare which
had to be produced by powerful electron beams allowing one to
explore the effects of various energy loss mechanisms on HXR and
MW emissions; and 3) it was located far from the disk centre that
allows us to compare the energy loss effects in a flaring
atmosphere seen by an observer from the Earth at larger viewing
angles.

The first detailed analysis of a spatial dynamics of this flare in HXR
and MW emissions was made by \citet{chandra2006}. They showed that
the spatial structure derived from the microwave emission allows
one to assume an interaction between two loops, one of which was a
small, newly emerging loop, while the other was a large overlying
loop, similar to the events identified by Hanaoka
\citep{hanaoka96, hanaoka99a, hanaoka99b}.

Later \citet{altyntsev2008} also interpreted the high-frequency MW emission in this
event produced by electrons precipitating into a
homogeneous source or into two homogeneous sources located in two loop legs where the depth variations of  the parameters of high energy electrons, plasma density and magnetic field were neglected. The authors though concluded
that such electrons were rather anisotropic, or beamed, because
(i)~the microwave emission consisted of many short (the order of
seconds) broadband pulses supposedly marking quite a few
elementary acts of beam injection into the loop legs, (ii)~the
microwave emission was O-mode polarized at 17 GHz (optically thin
part of the spectrum) meaning that the observed MW emission was an
immediate outcome of the beam precipitation rather than the
radiative transfer effects, (iii)~type III-like drifting bursts
were observed at lower frequencies meaning that beam electrons
streamed to higher atmospheric levels.

The goal of the current research is to extend the investigation of
this flare to simultaneous interpretation of HXR and MW emissions and to
fit their observed energy distributions with the simulations
of electron precipitation  into a converging magnetic loop by using the  Fokker--Planck kinetic
approach taking into account collisional and Ohmic losses in the
electric field induced by well collimated beam electrons.
Observations are described in section~\ref{obs}, the
model and method of simulations are described in
section~\ref{model}, the results of the model fit to observations
are shown in section~\ref{result} and the conclusions are drawn in
section~\ref{conc}.

\section{Observations}
 \label{obs}
\subsection{Instrumentation}

We used microwave total flux records from Nobeyama Radio
Polarimeters \citep[NoRP;][]{torii, shibasaki, nakajima} at 1, 2,
3.75, 9.4, 17, 35, and 80 GHz. The time resolution of routine NoRP
data available at the NoRP Web site is 1~s, and for flares it is
0.1~s. Nobeyama Radioheliograph \citep{nakajima94} produces images
at 17 and 34 GHz. The imaging interval of the flare mode data
which we used in this work was 0.1~s.

The line-of-sight magnetograms have been produced with the
Michelson Doppler Imager \citep[MDI;][]{scherrer} on SOHO. The
full disk magnetograms with a resolution of 1.98$^{\prime \prime}$
re-calibrated in December, 2008 were obtained from the Solar Data
Information Centre at Stanford University, US.

Information about HXR emission was supplied by instruments on
Yohkoh satellite \citep{kosugi91b, kosugi91a}. Images of HXR
sources and fluxes with a high temporal resolution (0.5~s) were
obtained from Hard X-Ray Telescope (HXT) data. HXT carried out
observations in four spectral bands: L, 14--23 keV; M1, 23--33
keV; M2, 33--53 keV; H, 53--93 keV. The spectra were obtained from
data of the Hard X-ray Spectrometer (HXS), which was one sensor of
the Wide Band Spectrometer \citep[WBS;][]{yoshimori, sato}. HXS
data provided hard X-ray spectra in a wide energy range of 20--657
keV. The temporal resolution of HXS data was 4~s, i.e., slightly
lower than the temporal resolution of HXT.

\begin{figure}
\centering
\parbox{0.9\hsize}{
\resizebox{
 \hsize}{!}{\includegraphics[width=4cm]{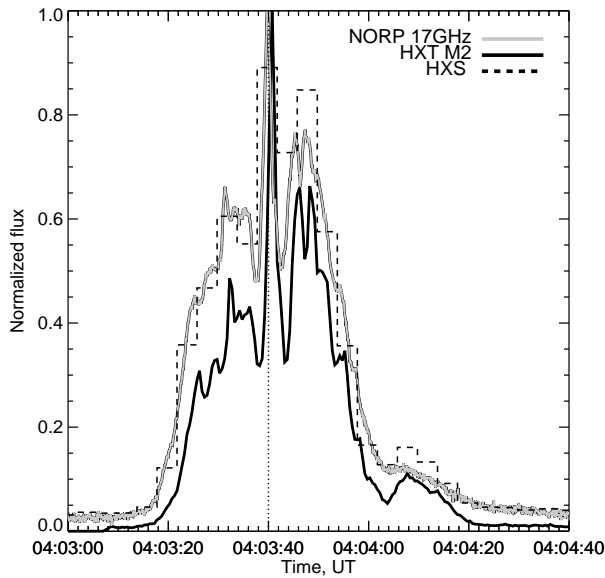}} }
 \caption{Light curves of the HXR burst registered in the HXT/M2 channel
(33--53~keV) and the HXS record integrated over the whole band
(20--657~keV, histogram) along with the microwave burst registered
by NoRP at 17 GHz. The vertical dotted line marks the main peak under consideration at
04:03:40 UT.}
  \label{f-wbs}
\end{figure}

\subsection{HXR and MW bursts and their sources}
 \label{observ}

The 1B/M6.7 flare of 10 March 2001 associated with a CME occurred
in active region 9368 (N27W42) and was previously studied in
detail in a number of papers listed above. However, recent
re-calibration of magnetic field strengths in SOHO/MDI
magnetograms alerted a review of previous results and inspired an
analysis of additional observations, which were not yet
considered.

\subsubsection{Light curves of MW and HXR emissions} \label{light}

As found by \citet{altyntsev2008}, broadband short pulses were
present in the whole frequency range 1--80 GHz of NoRP (see time
profiles in Fig.~\ref{f-norp}a). The MW time profiles in total
intensity at different frequencies were very similar. Three main
peaks are detectable in the MW burst with a total duration of
about 40~s (see Fig.~\ref{f-norp}a). The first MW peak at about
04:03:30.90 (\textit{all times hereafter are UT}) was the lowest
one, the second (central) MW peak at 04:03:40 with a total
duration of about 5~s was the highest one. The third MW peak had a
double structure with two sub-peaks occurring at 04:03:44.90 UT and
04:03:46.80 UT, respectively.

The time profiles of the polarized MW emission (Stokes V
component) around the central peak are shown in
Fig.~\ref{f-norp}b. They coincide with the light curves shown by
\citet{altyntsev2008}. The MW emissions of all the three peaks had
a positive (right-handed) circular polarization (RCP) at low
frequencies (1--3.75~GHz), and their polarization at higher
frequencies (9.4--35~GHz) was negative (left-handed, LCP), as
Fig.~\ref{f-norp}b shows.

In order to understand the dynamics of high-energy particles in
this flare, let us also use the whole energy range of the HXR
emission recorded by Yohkoh/HXS and estimate parameters of the
electron beam, thus extending studies carried out by
\citet{altyntsev2008} and \citet{chandra2006} that were confined
to the MW emission.

This flare was observed by Yohkoh/HXT in four energy channels L,
M1, M2, and H. Thus, the available energy bands of HXR channels
allowed us to derive photon spectra up to 600 keV. The HXS light
curve presenting the integrated flux within 20--657 keV range is
plotted along with HXT and NoRP data as shown in Fig.~\ref{f-wbs}.
The HXS light curve is plotted as a histogram to reveal its
temporal correspondence with the HXR and NoRP data. Sometimes the
HXS time bin could contain the emission integrated over two peaks
seen by HXT and at 17~GHz (for example, this happened at about
04:03:47). Fortunately, the temporal variations of HXR and NoRP
emissions during the main peak resemble the HXS light curve.

The time profiles of the HXR and 17~GHz emissions have a close
temporal correlation (Fig.~\ref{f-wbs}). Nevertheless, there are
some delays of HXR emission relative to microwaves. They amount to
a fraction of second at high frequencies and reach 1--2~s at low
frequencies \citep{altyntsev2008}.

\begin{figure}
\centering
\parbox{0.95\hsize}{

\resizebox{ \hsize}{!}{\includegraphics[width=\textwidth]{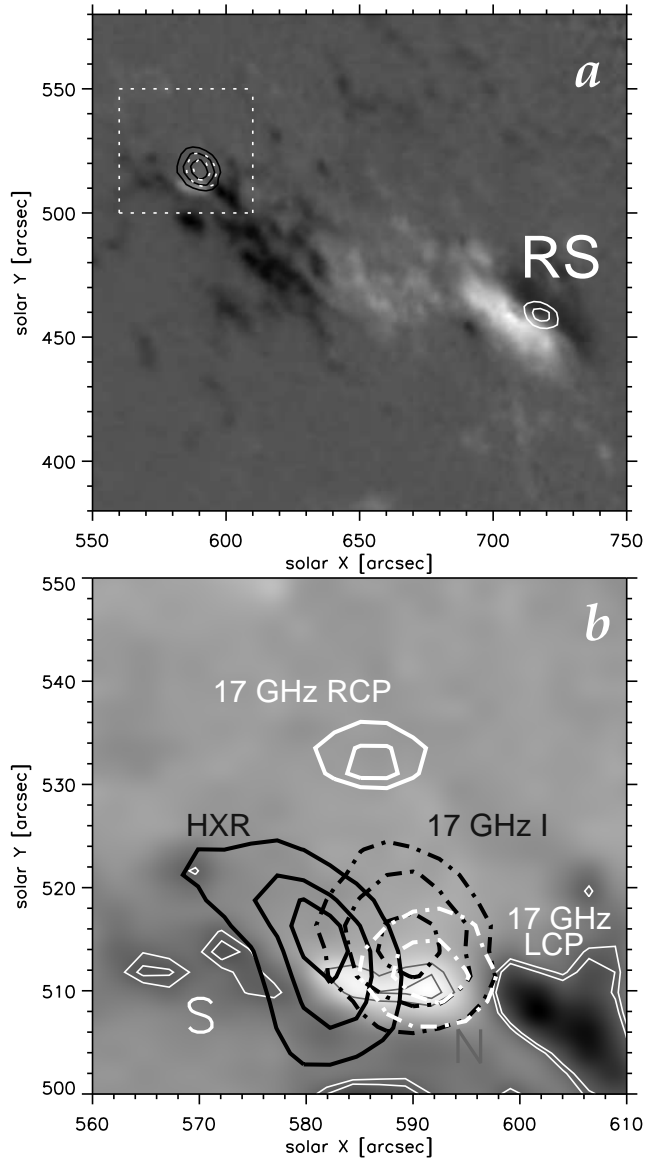}} }

\caption{a)~Contours of 17~GHz (NoRH, 04:04) flare sources
(black solid: Stokes I, 50\%, 70\% ,90\% of the maximum, white
solid: Stokes V, RCP, 70\%, 90\% of the maximum, white dotted:
Stokes V, LCP, 70\%, 90\% of the minimum) superimposed on an
MDI magnetogram (04:48:01.61, light areas represent N-polarity,
dark areas show S-polarity). The axes show hereafter arc seconds
from the solar disk centre. b)~Enlarged flare site denoted in
panel (a) by   the   broken frame.   Contours   of 17~GHz  (NoRH,
04:03:40) Contours levels are the same. Black thick contours
correspond to the HXR source (Yohkoh/HXT/H; 04:03:40.25; 20\%,
50\%, 80\% of the maximum). Thin solid gray contours mark
N-polarity [levels 600, 800 G], and thin white contours mark
S-polarity [levels $-600$, $-500$ G].}

   \label{F-overlaid}
 \end{figure}

\subsubsection{MW and HXR images}
 \label{image}

The flare was triggered by an interaction between two loops, one
of which was a small, newly emerging loop, while the other was a
large overlying loop \citep{chandra2006}. The two loops formed a
`three-legged' configuration in which the magnetic field had a
`bipolar + remote unipolar' structure. The footpoints of the small
loop (main flare source) can be seen in HXR/H and MW images
(Fig.~\ref{F-overlaid}b) whose position is denoted by the white
dotted frame in Fig.~\ref{F-overlaid}a. The magnetogram taken at
04:48:30 was compensated for the differential rotation to 04:03:40
(the most prominent peak time). The magnetic field strengths were
projection-corrected using the {\it zradialize} SolarSoftware
routine (Fig.~\ref{F-overlaid}). The length of the loop visible in
an EIT 195~\AA\ image is 10000 km according to
\citet{altyntsev2008}. During the flare, the NoRH beam size was
about $20.6^{\prime \prime} \times 17.3^{\prime \prime}$ at 17 GHz
and $9.5^{\prime \prime} \times 3.9^{\prime \prime}$ at 35 GHz.

The bulk of the MW emission at 17 GHz during the main peak was
generated from the small loop. At the onset of the burst an LCP
source located in the N-polarity magnetic field
(Fig.~\ref{F-overlaid}b) was observed. An RCP source appeared
north of the LCP source at 04:03:27 in the S-polarity magnetic
field, meaning that they both were polarized in the sense of the
O-mode.

The magnitude of the field in the spot labelled ``N'' (solid thin
gray contours in Fig.~\ref{F-overlaid}b) reached 800~G and for the
S-polarity field labelled ``S'' it reached 600~G. The large loop
had footpoints in the main and remote (RS) sources
(Fig.~\ref{F-overlaid}a). RS appeared at 04:03:51 and had a right
circular polarization at 17 GHz. The right polarization in the
main source disappeared after 04:03:52.

At the onset of the flare (04:03:12) a single compact source was
only seen in all energy channels of Yohkoh/HXT. HXR images
obtained with a 4-s cadence in the L channel showed that the
source expanded and got elongated by 04:04:05. In the H channel,
the compact source transformed to a loop-like one by 04:03:25, and
by 04:04:05 it became compact again. The distance between the
centroids of the radio and HXR sources was about 10$^{\prime
\prime}$.

For calculations of HXR spectra one needs areas of emitting
sources. We estimated the sizes of the sources obtained in all
four Yohkoh/HXT energy channels by fitting them with ellipses at
half magnitude for the positions as if these sources would be
located at the solar disk centre to exclude projection effects.
The estimated sizes of the HXR sources are $16^{\prime \prime}
\times 12.6^{\prime \prime}$, $18^{\prime \prime} \times
12.2^{\prime \prime}$, $15.4^{\prime \prime} \times 12^{\prime
\prime}$, and $12^{\prime \prime} \times 16.8^{\prime \prime}$ for
the L, M1, M2, and H channel, respectively.

\section{Model simulations}
 \label{model}
\subsection{Description of the model}
Let us consider the time-dependent Fokker--Planck approach for
precipitation of high-energy electrons injected into a cold
hydrogen plasma confined in a converging magnetic field structure
\citep{lea81,siversky09} while producing a self-induced electric
field, which forms a return current from the ambient plasma and
beam electrons \citep{kni77,ems80}. We take account of
collisions, Ohmic losses and pitch-angle anisotropy in scattering
on ambient particles \citep{dia88,mcc92a,zha10}.

There are the two basic expression to describe the problem: a
kinetic Fokker--Planck--Landau equation for beam electrons, and
the Ampere law for the neutralization of electric currents formed
from the direct and returning beams. A distribution function $f$
of beam electrons is governed by Fokker--Planck, or Landau,
equation \citep{lan37}:
\begin{eqnarray}
\frac{\partial f}{\partial t}+ V\cos\alpha\frac{\partial f}{\partial z}-
e\mathcal{E}V \cos\alpha\frac{\partial f}{\partial E}-
\frac{e\mathcal{E}\sin^2\alpha}{m_{\mathrm{e}}V}
\frac{\partial f}{\partial\cos\alpha}\nonumber\\
=\left(\frac{\partial f}{\partial t}\right)_{\mathrm{coll}}+
\left(\frac{\partial f}{\partial t}\right)_{\mathrm{magn}},\label{fp}
\end{eqnarray}
where $z$ is a linear depth measured from the top of a coronal
loop, $E$ and $\alpha$ are the electron energy and pitch angle,
respectively, $V$ is the electron speed, and $e$ and
$m_{\mathrm{e}}$ are the electron charge and mass, respectively.
The first term on the left-hand side describes variations of the
distribution function in time ($t$), the second term describes its
variations with depth ($z$), the third and the fourth terms
reflect energy losses owing to Ohmic heating and pitch-angular
diffusion under the presence of a self-induced electric field
$\mathcal{E}$. The two terms on the right-hand side $(\partial
f/\partial t)_{\mathrm{coll}}$ and $(\partial f/\partial
t)_{\mathrm{magn}}$ describe the energy of a particle and the
pitch-angle diffusion caused by scattering on ambient particles
(collisional integral) and in a converging magnetic field,
respectively.

The collisional integral is taken in the linearized form suggested
by \citet{dia88}, considering  variations of collisional energy losses and anisotropic scattering with pitch angle diffusion for a given kinetic temperature of the ambient plasma and distribution function $f$ of beam electrons as follows

\begin{eqnarray}
\left(\frac{\partial f}{\partial t}\right)_{\mathrm{coll}}=
\frac{1}{\varv^2}\frac{\partial}{\partial\varv}\left[\varv^2\nu(E)
\left(\frac{k_{\mathrm{B}}T_{\mathrm{e}}}{m_{\mathrm{e}}}
\frac{\partial f}{\partial\varv}+\varv f\right)\right]\nonumber\\
+\nu(E)\frac{\partial}{\partial\cos\alpha}\left(\sin^2\alpha
\frac{\partial f}{\partial\cos\alpha}\right).\label{collint}
\end{eqnarray}

Here $k_{\mathrm{B}}$ is the Boltzmann gas constant, $T_{\mathrm{e}} (\xi)$ is the
ambient  electron temperature at column depth $\xi$ and $\nu(E)$ is the frequency of collisions
taken as follows

\begin{displaymath}
\nu(E)=\frac {4k}{3} \frac{\sqrt{2\pi}}{m_{\mathrm{e}}} n e^4\lambda(\xi)\,E^{-3/2},
\end{displaymath}

where $n$ is the ambient plasma density, $E$ is the beam electron
energy, $ \lambda(\xi)$ is the Coulomb logarithm for collisions of
beam electrons with the ambient ones for a column depth $\xi$, and
the parameter $k$ is taken in the form \citep{ems78}

\begin{displaymath}
k= 2x\,+\,(1-x) \frac{\lambda ^{\prime \prime} - \lambda^{\prime}}{\lambda},
\end{displaymath}
where $\lambda ^{\prime \prime}$ and $\lambda^{\prime}$ are the Coloumb logarithms for collisions of beam electrons with the ambient ions and neutrals, respectively.

The linear coordinate $z$ is replaced by the column density$\xi(z)=\int\limits_{z_{\min}}^zn(s)\,\mathrm{d}s$\citep{dia88,siversky09}. The effect of the magnetic field convergence variations with column density is described by the semi-empirical expression \citet{siversky09} with the characteristic column depth of $10^{20}$ $cm^{-2}$, allowing to match the observations of magnetic field magnitudes at the photosphere and chromosphere \citep{kontar08}.
This semi-empirical form was further modified by \citet{siversky09} for the  magnetic field distribution with a linear depth, required for the simulation of MW emission.


\subsubsection{Self-induced electric field}
The problem of electrostatic electric field carried by beam electrons and the return current formed by the ambient plasma was discussed in the past decades by many authors \citep[see, for example][and the references therein]{Knight1977,ems80}.  \citet{brown1984aa} concluded that a return current can be established electro-statically because of large radii of electron beams inside flaring atmospheres that implies very long resistive timescales along the finite beam lengths leading to negligible inductive effects compared to the electrostatic ones even for plasma with anomalous resistivity. The full electro-magnetic approach carried out by \citet{oord90} to study of electric and magnetic fields induced by precipitating electron beams and their interaction with electrostatic field carried by these electrons allowed to conclude that the both electric fields: electrostatic and induced ones co-exist in a flaring atmosphere and are neutralised by the electrostatic or solenoidal response of the ambient plasma.

These two plasma responses are found to act independently \citep{brown1984aa,oord90} forming from one side a return current from the ambient electrons moving in the opposite direction to precipitating beam which neutralises the electrostatic electric field \citep{dia88,zha95,zha06,bat2008}, and  from the other side by ambient electrons compensating for solenoidal electric (and magnetic) fields which can be often observed as longitudinal transient magnetic fields \citep{2001ApJ...550L.105K,Zharkova2005JGR,2005ApJ...635..647S}. Therefore, for the purpose of this study, similarly to \citet{dia88,zha06}, we consider only electrostatic part of the electric field induced by beam electrons and
 assume that the electric current carried by beam electrons in the ambient plasma serving as the conducting media is compensated by the return current formed from the ambient electrons.  This allows us to estimate a magnitude of the electro-static electric field induced by the electron beam in the ambient plasma as follows \citep{dia88,zha95}:
\begin{equation}
 \label{rc}
\mathcal{E(\xi)}=\frac{j(\xi)}{\sigma(\xi)}=
\frac{e}{\sigma(\xi)}\int\limits_{E_{\min}}^{E_{\max}}V(E)\,\mathrm{d}E
\int\limits_{-1}^1 f(\xi, E, \mu)\mu\,\mathrm{d}\mu,
\end{equation}
where $\sigma(\xi)$ is the classical plasma conductivity at a
given temperature of the ambient plasma defined as (Spicer, 1977):

\begin{eqnarray}
\frac{1}{\sigma}&=&\frac{7.28\times 10^{-8}X}{T_{\mathrm{e}}^{1.5}}
\ln\left(\frac{3}{2e^3}\frac{k_B^3T_{\mathrm{e}}^3}{\pi n}\right)\nonumber\\
&+&\frac{7.6\times 10^{-18}(1-X)}{X}T_{\mathrm{e}}^{0.5},\label{spcond}
\end{eqnarray}

where $X$ is the ionization degree of the ambient plasma.

The electric field induced by beam electrons carries a dual role:
firstly, it induces (within the timescale of the double collisional
time) the electro-magnetic field in the ambient plasma
\citep{oord90} and sets up the return current in the ambient
plasma, and secondly, turns the precipitating beam electrons to
the direction opposite to their original one, returning them
back to the source and, thus, reducing the magnitude of the return
current from the ambient electrons by the factor of two or higher
\citep{zha10}.

\subsubsection{The initial and boundary conditions}
It is assumed that at the moment of injection there are no beam
electrons in the atmosphere, i.e. $f(E,\mu,s,0)\,=\,0 $. The
distribution function of beam electrons on the top boundary
$\xi=\xi_{\min}$ is assumed to have power law distribution
in energy in a range
from $E_{\min}$ to $E_{\max}$ with a spectral index $\gamma+0.5$ (that corresponds to the particle energy flux varying with energy as a power-law with the index $\gamma$ \citep{Syrovatsky72}) and normal distribution in the cosines $\mu$ of  pitch-angle $\alpha$
($\mu=\cos\alpha$) with the half-width dispersion $\Delta\mu$  as
defined below:

\begin{equation}
\left.f(E, \mu, t)\right|_{\xi=\xi_{\min}}=\left\{\begin{array}{l}
\displaystyle
AE^{-\gamma-0.5}\exp\left[-\frac{(\mu-1)^2}{\Delta\mu^2}\right]U(t),\\[10pt]
\textrm{for}~E_{\min}\le E\le E_{\max},~\mu>0,~t\ge 0,\\[10pt]
0,~\textrm{elsewhere}.
\end{array}\right.
\label{mu_dist}
\end{equation}

This condition represents an electron beam injected with a
power-law dependence in energy and a narrow dispersion
($\Delta\mu\ll 1$) in the pitch angle centered at $\alpha=0$ (or
$\mu=1$). $U(t)$ is a temporal profile of the electron beam
injection, which denotes the initial beam flux variations during a
required time interval, accepted to be equal to unity (a steady injection).

The coefficient $A$ is found from the normalization condition for an
electron distribution function as follows:
\begin{equation}\label{norm}
A \,=\,
\frac{F_{0}}{\int\limits_{E_{\min}}^{E_{\max}}E^{1/2}\mathrm{d}E\int\limits_{-1}^1
f(E, \mu, \xi_{\min})\,\mathrm{d}\mu},
\end{equation}
where $F_{0}$ is the initial energy flux of accelerated electrons
on the top boundary, normally derived from observations and set up
as a free parameter in simulations.

\subsubsection{Method of solution and accepted parameters}
 \label{method}

The set of equations (\ref{fp}) and (\ref{rc}) defines the
electron beam distributions in a flaring atmosphere at every
instant of precipitation. For their solution we use the summary
approximation method described by \citet{siversky09} splitting the
time interval into three bits and solving simultaneously the
temporal equations for electron distribution changes in depth,
energy and pitch angle cosines. The solutions are sought for a
electrons with relativistic correction precipitating in a quasi-stationary regime,
 e.g. when the electron distributions in depth and pitch angles do not change in
time. This is normally achieved within
0.07--0.2~s after the beam onset when the flows or precipitating
and returning electrons form a steady circuit \citep{siversky09}.

The simulations are carried out for electrons in the energy range
from 12 keV to 1.2 MeV. Electron distribution functions at higher
energies (up to 10 MeV) were obtained by extrapolation of the
mentioned simulations by using power-law fit. Comparison with the
results of simulations covering a wider energy range has shown
that such an interpolation provides very high accuracy, while
greatly reducing computation time for the relativistic  FP
equation.

The following precipitation models are used in our simulations:
model C including only Coulomb collisions with the charged
particles and neutrals of the ambient plasma; model C+E including
the collisions and Ohmic losses in a self-induced electric field;
model C+B including the collisions and scattering in converging
magnetic field and model C+E+B considering all the above factors
(collisions, electric field and magnetic field convergence).

\begin{figure}
\includegraphics{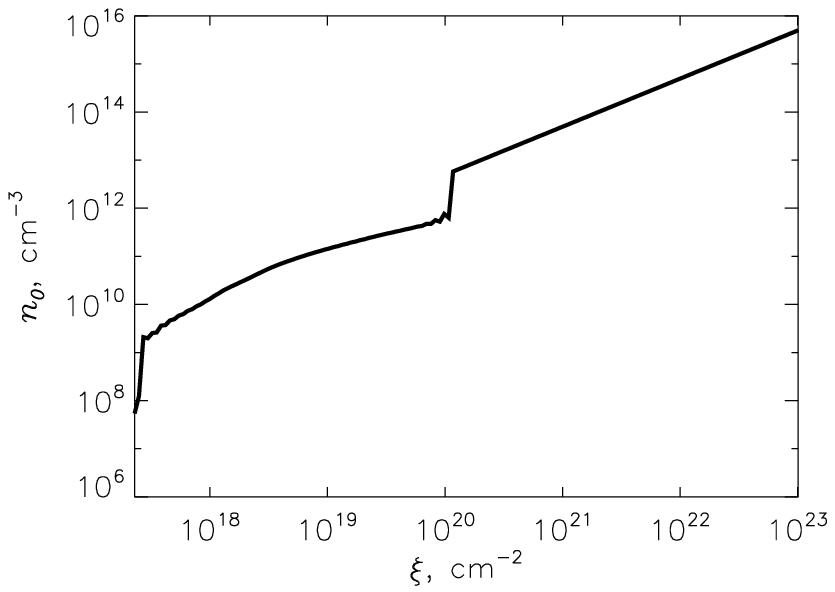}
\includegraphics{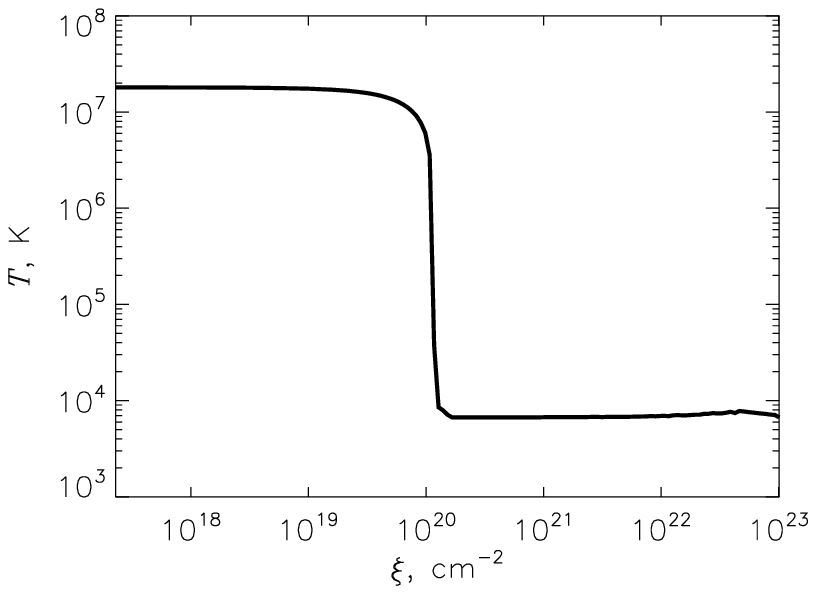}
\caption{Height profiles of the plasma parameters (electron density and temperature vs. column density) which are used in the simulations.}
\label{FigHDprofiles}
\end{figure}

In a flaring loop, both the electron density and temperature depend on height. These dependencies affect the plasma conductivity (\ref{spcond}) and collisional integral (\ref{collint}) and thus must be taken into account. In this work, we use the temperature and density profiles derived from the hydrodynamic simulations \citep{Zharkova07} shown in Fig. \ref{FigHDprofiles}. We assume that the boundary of the transition region is located at the depth where the column density $\xi$ equals $10^{20}$ $\textrm{cm}^{-2}$, and the profiles demonstrate a sharp change of the plasma parameters around this depth.

\section{Results of simulations}
 \label{result}
\subsection{Estimation of beam parameters}
\label{estim}

\citet{altyntsev2008} evaluated some plasma parameters from
observations, so that the peak injection rate of electrons above
10 keV was found to be $8.5 \times 10^{36}$ electrons s$^{-1}$, the total
number of emitting electrons in the radio source was $3.83 \times
10^{36}$ electrons, the background plasma density was  $3 \times
10^{11} \textrm{cm}^{-3}$, the single power-law electron spectrum
index $\gamma$ was 2.4, angular scale of the MW source was
6$^{\prime \prime}$, the magnetic field value at the photosphere
near the footpoints of the flare loop was 170--340 G.

The authors \citep{altyntsev2008} received the best fit for the
following beam parameters: the characteristic lifetime of the
emitting electrons in the radio source $\tau_\mathrm{l}$=0.45 s,
viewing angle between the line of sight and the direction of the
magnetic field at the source $\theta$=$80^{\circ}$,
a beam precipitating along the direction of $\mu_0=0.5$ with the pitch angle dispersion at the half-width corresponding to $\Delta\mu=0.35$.

The current estimations of the HXR emission observed by Yohkoh (see section \ref{observ}) gave the total energy fluxabout $1.5 \times10^{11}$~erg/s/cm$^{2}$ for energies from 10 keV
to 100 keV and $\gamma=$ 2.4 that was very close to the one
estimated by \citet{altyntsev2008}.  However, we extended the HXR photon energy range
to 200 keV and for this energy range (80-200 keV) we derived the spectral index to be $\gamma=$ 3.12. Hence, we have at hands the double power law photon spectrum with a
break energy  about 80 keV.

As shown by \citet{zha06,siversky09} such double law HXR photon energy spectra are formed by combined collisional and Ohmic energy losses during precipitation of a single power law electron beam with a  spectral index of beam electrons equal to that of the upper photon energy range (80--200 keV).  Ohmic losses lead to flattening of he HXR photon spectrum at lower energies, which is higher for beams with higher initial energy fluxes and spectral indices.

Also it was shown \citep{zha06,Zharkova11} that, in order to calculate a total energy flux  of beam electrons accountable for the observed HXR photons, one needs to prolong the HXR photon spectrum at higher energy (with index 3) to a lower energy range, to account for a single power law electron beam causing this emission. Since the  higher energy spectral index is higher than the one calculated with the spectral index (2.3)derived at the  lower energy range we obtain the new, enhanced, total energy flux  F = $10^{12}$ erg/s/cm$^{2}$ which accounts for both collisional and Ohmic energy losses by beam electrons.

This difference in spectral indices and energy fluxes for lower
and higher energy bands resembles very closely the HXR photon
spectra produced in the models of electron beam precipitation with
the self-induced electric field which causes their flattening
towards lower energies \citep{zha06}. This indicates a need to
consider the model of electron beam precipitation taking into account Ohmic
losses, in addition to collisions, and for the beam
parameters  to adopt those derived for the higher energy range.

In addition, because of the announcement in 2008 by MDI team about the problems with the magnetic field calibration leading to twice or smaller magnitudes of the measured magnetic field with the previous one, we needed to  correct the measured magnetic field in the
footpoints.  With new calibration the magnetic field magnitudes at the photosphere are increased to 600--800 G compared to the magnitudes of 170--340 G accepted in the previous
study by \citet{altyntsev2008} [the importance of the correction
of the magnetic field strength to reconcile MW and HXR spectra was
recently demonstrated by \citet{kun09}]. Also, the calculations of
HXR and MW emission in the current paper are carried out for the
the visible HXR source area of about 150 arcsec$^{2}$ (see
section~\ref{observ}). The fact that in this flare there is a presence  of broadband MW emission in
a wide frequency range (1--80 GHz)  simultaneously with HXR emission from 10 to 200 keV allows to assume that  they both are produced by the same populations of electrons, one needs to accept the energy  for beam electrons ranging from 12~keV to 10~MeV.

Another improvement of the previous simulation model was related
to consideration of a flaring atmosphere highly inhomogeneous in atmospheric depth that was contrary to the homogeneous models of both ambient and beam electrons
considered in each of the two footpoint sources by \citet{altyntsev2008}.
The density and temperature variations are derived as a result
of hydrodynamic response to the heating caused by injection of beam
electrons \citep{Zharkova07}. This model produces significant depth
variations not only in the physical conditions in the ambient
plasma but also of the beam electrons themselves if different energy loss and pitch angle scattering mechanisms are considered.

\subsection{Simulated HXR and MW emission}
 \label{emission}

Both HXR and MW emissions are assumed to be produced by the same
population of power law beam electrons with a spectral index of 3
ranging energies from 12 keV to 10 MeV as derived in
section~\ref{estim}. These beam electrons are steadily injected into a flaring
atmosphere along the pitch angle zero ($cos \mu=1$) with a normal distribution are getting scattered by the ambient particles, self-induced electric and  converging magnetic field.
The electron distribution functions for different models of energy losses (C,
CE, CB and CEB) are discussed in detail in \citet{siversky09,zha10,kuz10}. This defines the electron beam dynamics
and pitch-angle distributions at every precipitation depth.

The main issue in electron distributions is a formation of returning
electrons either by magnetic field convergence or by self-induced
electric field of precipitating beam electrons. These two beams,
precipitating and returning ones, are producing either HXR
emission in scattering on the ambient plasma particles or slowing down
in the electric field and MW gyro-synchrotron emission from their
gyration in a strong converging magnetic field.  Electrons emit HXR photons in the directions downwards (to the photosphere with pitch angle $\alpha=0-90^\circ$) and upwards (to the observer at the Earth, with $\alpha= 90-180^\circ$) if the loop stands near the solar disk centre. If the loop is located outside the central zone, the viewing angles for HXR and MW observations will be amended by the angle of the loop position (latitude and longitude) on the solar disk which described in detail by \citet{zha10}.

\subsubsection{HXR emission and directivity}
 \label{hxr_em}

We calculate photon spectra of HXR emission integrated over all atmospheric depth, polarization and directivity normalized on the average intensity over all angles for the
relativistic cross-sections with pitch angle and viewing angle
dependence as described below.

\begin{eqnarray}
\left[\begin{array}{c}
I\\
Q
\end{array}\right](h\nu, \theta)&=&
\frac{A}{2\pi\mathcal{R}^2}\int\limits_0^{\xi_{\max}}\mathrm{d}\xi
\int\limits_{h\nu}^{\infty}\varv(E)\,\mathrm{d}E\int\limits_{-1}^1
f(\xi, E, \mu)\,\mathrm{d}\mu\nonumber\\
&\times&\int\limits_0^{2\pi}
\left[\begin{array}{c}
\sigma_I\\
\sigma_Q
\end{array}\right](h\nu, \theta, E, \mu, \varphi)\,\mathrm{d}\varphi,
\label{Xint}
\end{eqnarray}
where $A$ is the cross-section  area of a magnetic tube, R is the astronomical unit, $I$ and $Q$ are the  Stokes parameters related to the intensity and linear polarization of the emission.

Then a degree of linear polarization can be defined as follows
\begin{equation}
\eta=\frac{Q}{I},
\label{pol}
\end{equation}
and the distribution of any emission in the angles $\theta$ can be described by a
directivity
\begin{equation}
D(\theta)=\frac{I(\theta)}{\left<I\right>},
\label{dir}
\end{equation}
where $\left<I\right>$ is the emission intensity averaged over all the angles. The three-dimensional integrals over the electron velocity in (\ref{Xint}) are calculated with Monte-Carlo method.

\begin{figure}[h]
\centering
\parbox{0.7\hsize}{
\resizebox{\hsize}{!}{\includegraphics{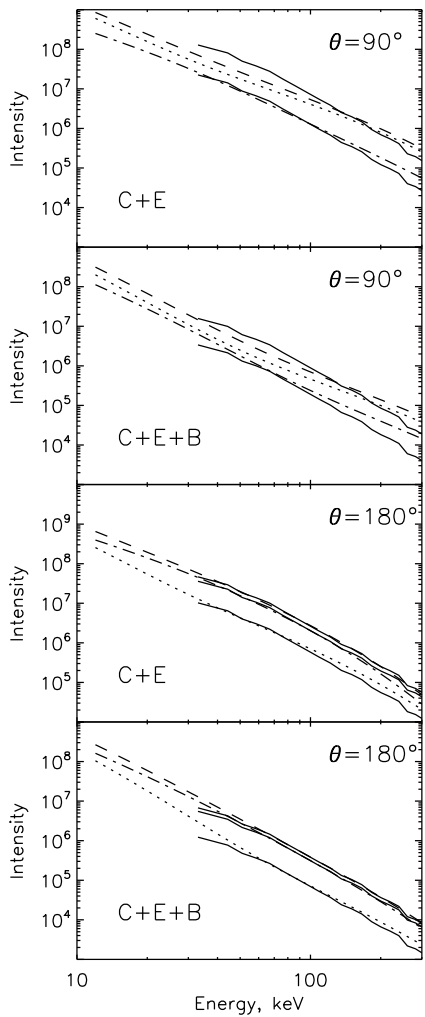}} }
 \caption{Intensity (in arbitrary units) of the HXR emission spectra
calculated from equation (\ref{Xint}) for different energies,
models and propagation directions (0 to $90^\circ$ for downward
emission and from $90^\circ$ to $180^\circ$ for upward emission).
Dash-dotted line: emission from upward propagating particles;
dotted line: from downward propagating particles, dashed line:
total emission (downward + upward); solid line: observational data
multiplied by the corresponding factors to match the simulated
absolute intensities.} \label{HXR-model}
 \end{figure}

The simulations of HXR photon spectra for different
precipitations models (C, CB, CE and CEB) including those emitted
downwards (to the photosphere) and upwards (to the observer) and
for different magnitudes of viewing angles are plotted in
Fig.~\ref{HXR-model} from the top to the bottom, respectively. On
top of the simulated energy spectra we over-plotted by a solid
line the energy spectrum deduced from the observations (see
section~\ref{observ}).

The plots of HXR emission simulated for different directions of electron
propagation (downwards with pitch angle cosines $\mu > 0$ and
upwards with $\mu < 0$) reveals that in the corona the majority of
electrons moves in the downward direction while only a twice
smaller number moves upwards. In the chromosphere, it is to the
contrary: most electrons move upwards and only a smaller fraction
of them keeps moving downwards. Thus, for this flare both effects:
albedo and Ohmic losses have to be very significant at upper
(coronal) precipitation depths.

The spectra calculated for CB model (collisions and converging magnetic field) show
the highest intensity compared to CE and CEB models \citep{zha10}. This is
because the electric field induced by precipitating electron beam prevents the
particles from reaching deeper layers where the bulk of electrons
(with lower cutoff energy) can emit HXR photons,  thus reducing
the intensity. The CE model provides a smaller intensity than for
model CB while the intensities for CEB model are much smaller than
the either of the above for CB or CE. This happens because of the
combined effect of the two factors (convergence and electric
fields) which significantly reduces a number of particles being able to
precipitate into deeper atmospheric layers where the bulk of
electrons can emit HXR photons. \citet{zha10} have shown that for
$F=10^{10}$ erg/s/cm$^{2}$, the results were opposite: the CE
model provides a higher HXR intensity than the CB model. This confirms
the conclusion by \citep{zha10} that for $F=10^{12}$
erg/s/cm$^{2}$ the effect of a self-induced electric field is much
stronger than for a weaker beam.

It can be also noted that the observed HXR photon spectrum has the best
fit by the HXR emission emitted upwards plus downwards (the full albedo effect) under a viewing angle of $180^\circ$ by an electron beam with given parameters for the CE (collisions and electric field) and CEB (collisions, electric field and converging magnetic field with convergence equal to three) precipitation models. The HXR
spectrum observed in this flare reveals a noticeable flattening towards lower
energies which is better reproduced by CE model indicating a significant
effect of the self-induced electric field in the energy losses by beam
electrons. The observed and simulated energy
spectra reveal the best fit for the viewing angle of
$180^{\circ}$ (the direction towards the observer looking from the
top). Given the flare location at the location (N27W42) one can
assume that the flaring loop emitting HXR photons is tilted towards the observer by $10^\circ-40^\circ$.

  \begin{figure*}    
  \centerline{\hspace*{0.015\textwidth}
               \includegraphics[width=0.455\textwidth,clip=,
                              bb=0 30 283 226]{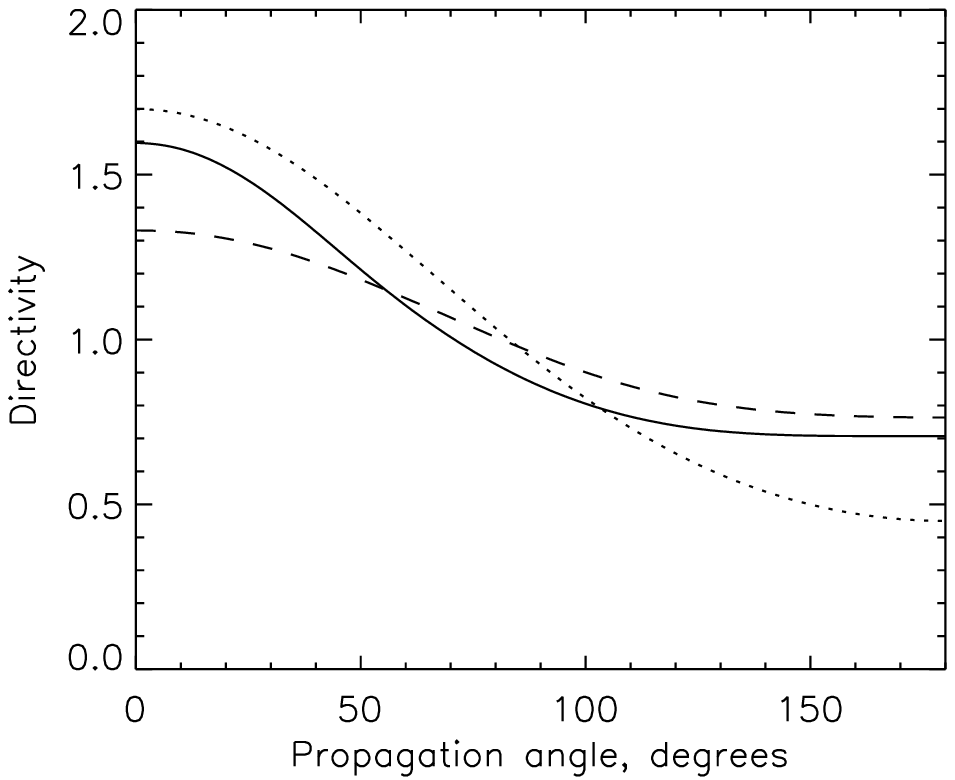}
               \hspace*{-0.03\textwidth}
               \includegraphics[width=0.455\textwidth,clip=,
                              bb=0 30 283 226]{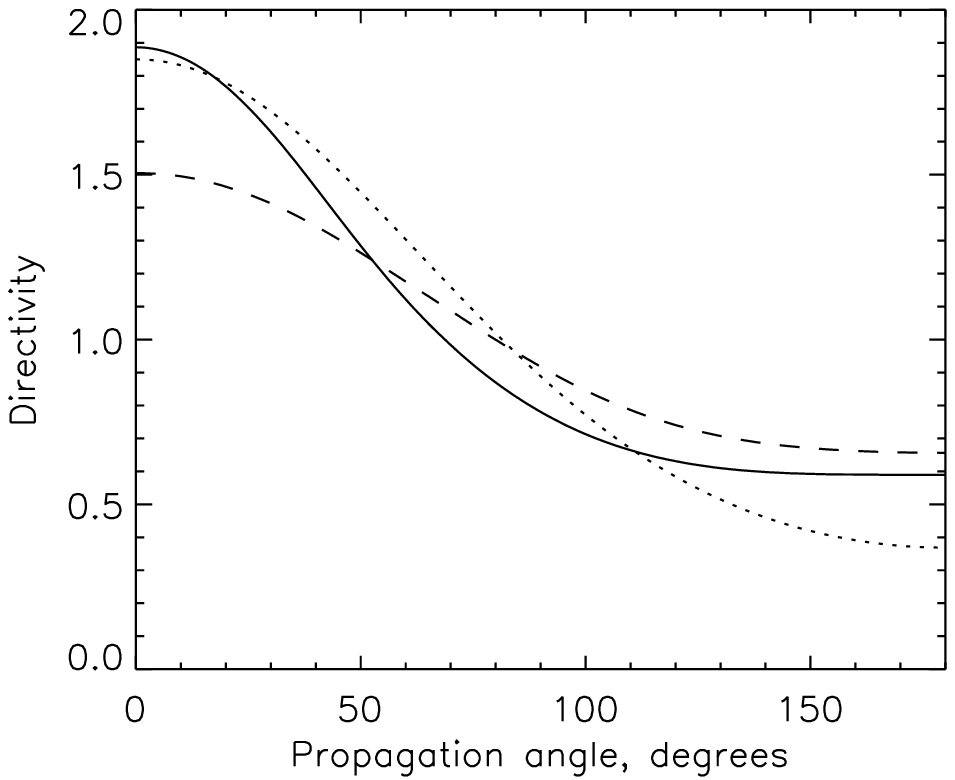}
              }

     \vspace{-0.30\textwidth}   
          \centerline{\large      
      \hspace{0.29 \textwidth}  {30 keV}
      \hspace{0.37 \textwidth} {50 keV}
         \hfill}
     \vspace{0.275\textwidth}    

   \centerline{\hspace*{0.015\textwidth}
               \includegraphics[width=0.455\textwidth,clip=]{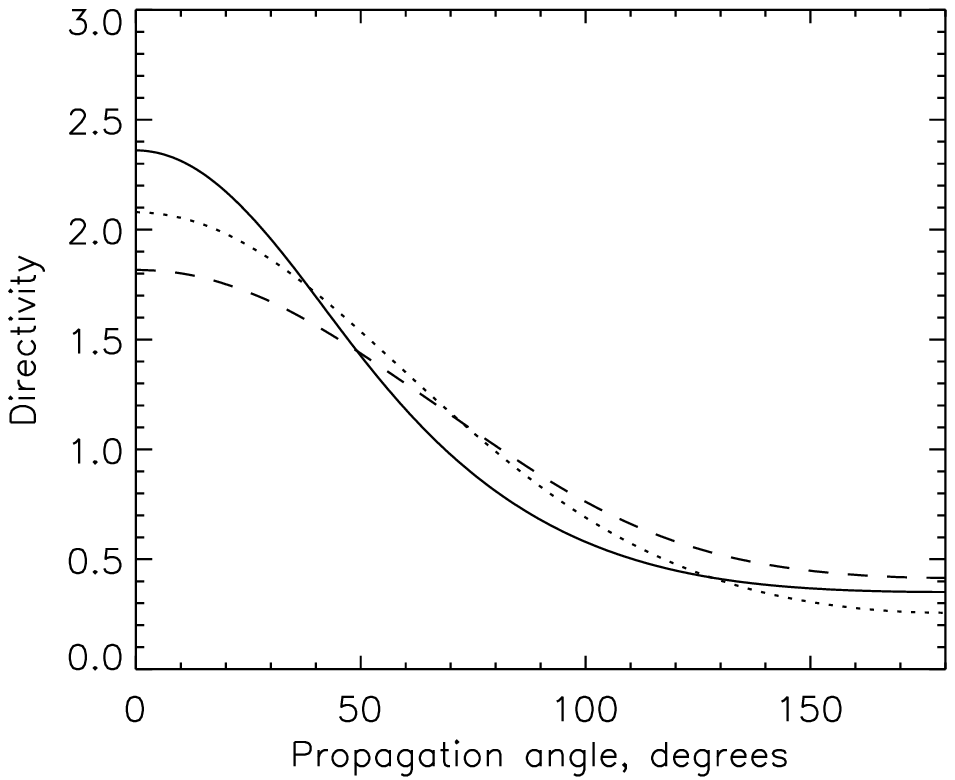}
               \hspace*{-0.03\textwidth}
               \includegraphics[width=0.455\textwidth,clip=]{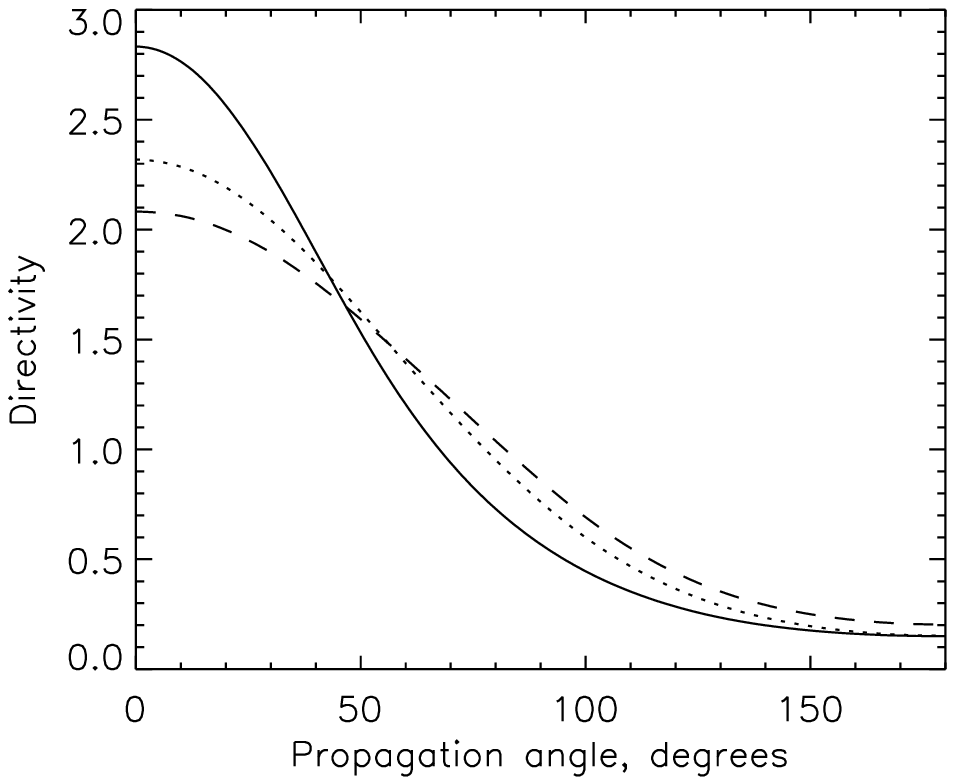}
              }

     \vspace{-0.35\textwidth}   
          \centerline{\large      
      \hspace{0.29 \textwidth}  {100 keV}
      \hspace{0.37 \textwidth} {200 keV}
         \hfill}
     \vspace{0.33\textwidth}    

 \caption{Directivity of the HXR emission calculated from equation (\ref{dir}) for electron energies of 30 keV,
50 keV, 100 keV and 200 keV for different models of electron
energy losses and propagation directions. Solid line for CE model,
dotted line for CB and dashed line for CEB model. Propagation
direction indicates the viewing angle for the observer looking
from the top and varies from 0 to $90^\circ$ for downward emission
and from $90^\circ$ to $180^\circ$ for upward emission. }
   \label{HXR-dir}
\end{figure*}

At the same time the directivity of HXR emission is found to range
between 2 and 3 for the electrons propagating downwards and about
0.5 for the electrons moving upwards. This points out to clear
anisotropy of the electron beam producing this HXR emission with
domination of the downward moving electrons. In order to see this
emission from the top where the observer is placed in our models,
HXR photons have to be reflected by the photosphere, e.g. the photospheric albedo effect should play a significant role
\citet{kon06}. In the present study we have assumed 100$\%$ albedo coefficient for the emission emitted downwards while in
reality this can have more complicated dependence on energy of electrons \citep{kon06} and their pitch-angle distribution. However, this effect requires to consider strong electron anisotropy which we described above, contrary to the isotropic electron distributions considered by \citet{kon06} that will be considered in the forthcoming paper.

\subsubsection{MW emission}
 \label{mw_em}

We now calculate the parameters of MW emission. The model of the emission
source is similar to that used by \citet{kuz10}: we assume that the
parameters of the coronal magnetic tube (such as the plasma density,
magnetic field, and the parameters of the accelerated electrons) depend
only on the coordinate $z$, a linear distance along the tube.
Also, we used a stratification of MW emission across the layers in vertical direction (corresponding to different heights in the atmosphere). These layers are assumed to be quasi-homogeneous sources, so that the emissivity and absorption coefficients within each layer are assumed to be constant. As result, the total MW emission emitted from the whole atmosphere is equal to a sum of the contributions by all layers written as follows:

\begin{equation}
I_{\sigma}=\frac{D}{\mathcal{R}^2}\int\limits_0^{z_{\max}}
\frac{j_{\sigma}(z)}{\varkappa_{\sigma}(z)}
\left[1-e^{-\varkappa_{\sigma}(z)L}\right]\,\mathrm{d}z.
\label{mwint}
\end{equation}

Here $I_{\sigma}$ is the emission intensity (observed at the
Earth) of the magneto-ionic mode $\sigma$, $D$ and $L$ are the
visible diameter of the magnetic tube and the source depth along
the line-of-sight, respectively, and $\mathcal{R}$ is the
astronomical unit. The equations for the gyrosynchrotron plasma
emissivity $j_{\sigma}$ and absorption coefficient
$\varkappa_{\sigma}$ are given by e.g. \citet{melrose68} and
\citet{ramaty69}. The polarization degree is defined as

\begin{equation}
\eta=\frac{I_{\mathrm{X}}-I_{\mathrm{O}}}{I_{\mathrm{X}}+I_{\mathrm{O}}},
\label{mwpol}
\end{equation}

where $I_{\mathrm{O}}$ and $I_{\mathrm{X}}$ are the intensities of
the ordinary and extraordinary modes, respectively. The
directivity of MW emission is calculated for different propagation
(or viewing) angles), similar to that of HXR emission (see
equation \ref{dir}) with the intensity calculated by using
equation (\ref{mwint}).

Unlike the previous paper \citep{kuz10}, in the present study we
use the temperature and density of a flaring model derived from the
hydrodynamic simulations \citep{Zharkova07} with the distance $z$ from the
injection point varying from 0 to $z_{\max}\simeq 10\,000$ km, the thermal
plasma density varying from $2\times 10^9$ $\textrm{cm}^{-3}$ to $2\times
10^{13}$ $\textrm{cm}^{-3}$ (see Fig. \ref{FigHDprofiles}), and the magnetic field strength of
$B=780$ G at the characteristic column depth of $\xi_c=10^{20}$
$\textrm{cm}^{-2}$ (or $z=z_{\max}$). The parameter $z_{\max}$ corresponds to the boundary of the transition region, since only the coronal part of the loop makes a contribution into the MW emission (in the deeper layers, the plasma density is too high).

In the models with converging
magnetic field, the field strength varies with depth from the top of the
emission source to this characteristic depth \citep{siversky09}, which is
another factor (in addition to the variations of a distribution function)
affecting the MW emission parameters; we used the models with the
convergence factors $B_{\mathrm{footpoint}}/B_{\mathrm{top}}=2$ and 3
which provided the very reasonable fit to the observations (see sections \ref{hxr_obs} and \ref{mw_obs}). In
the CE model, variation of the magnetic field with height (with the
convergence factor 3) was considered when calculating the plasma
emissivity and absorption coefficient, but the effect of converging
magnetic field on the electron distribution was neglected. Also, we assume
that the loop width is $D=10\,000$ km and the source depth along
line-of-sight is $L=10\,000$ km; these parameters agree with the imaging
observations and provide the best agreement of the calculated MW spectra
with the observed ones (see below).

\begin{figure*}
 \centering
   \includegraphics[width=6cm]{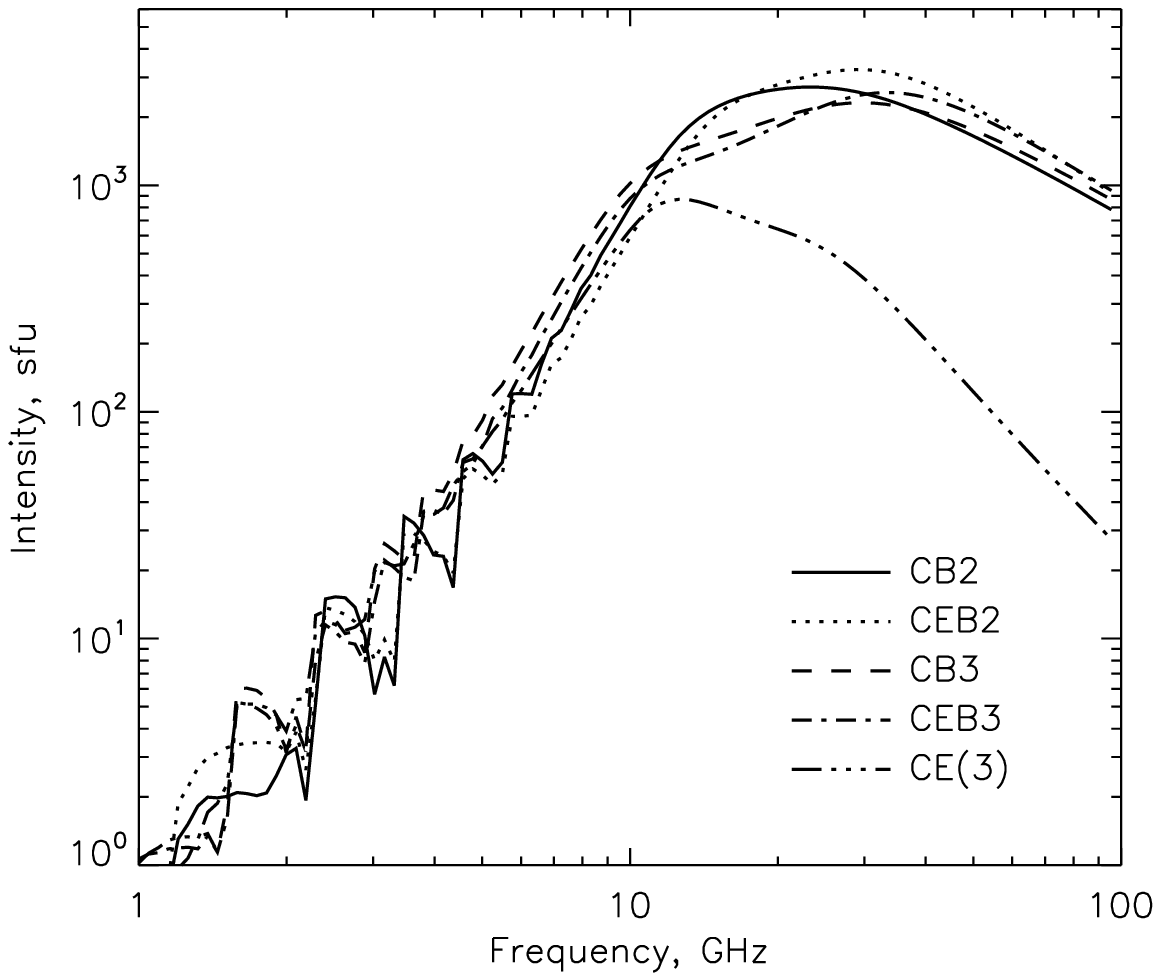}
   \includegraphics[width=6cm]{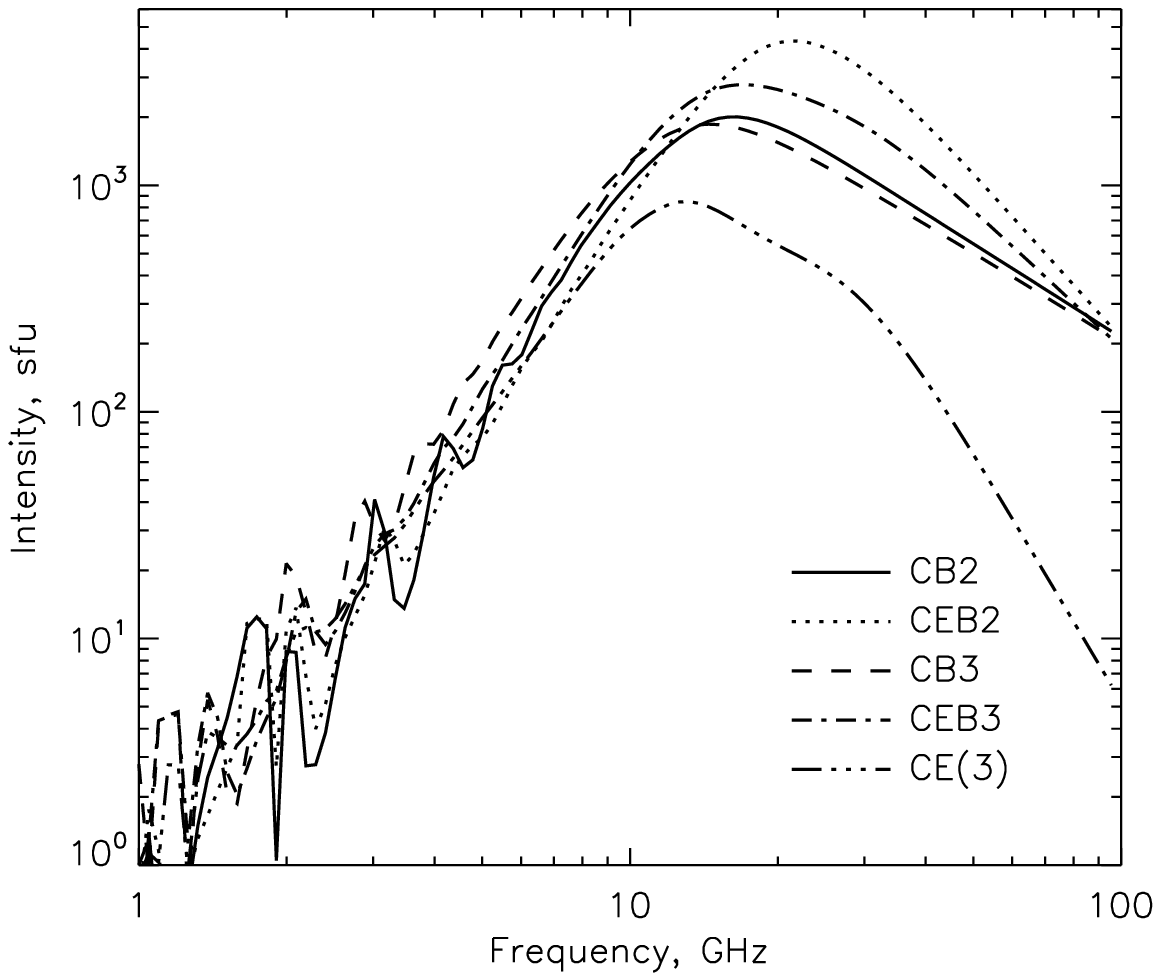}
   \includegraphics[width=6cm]{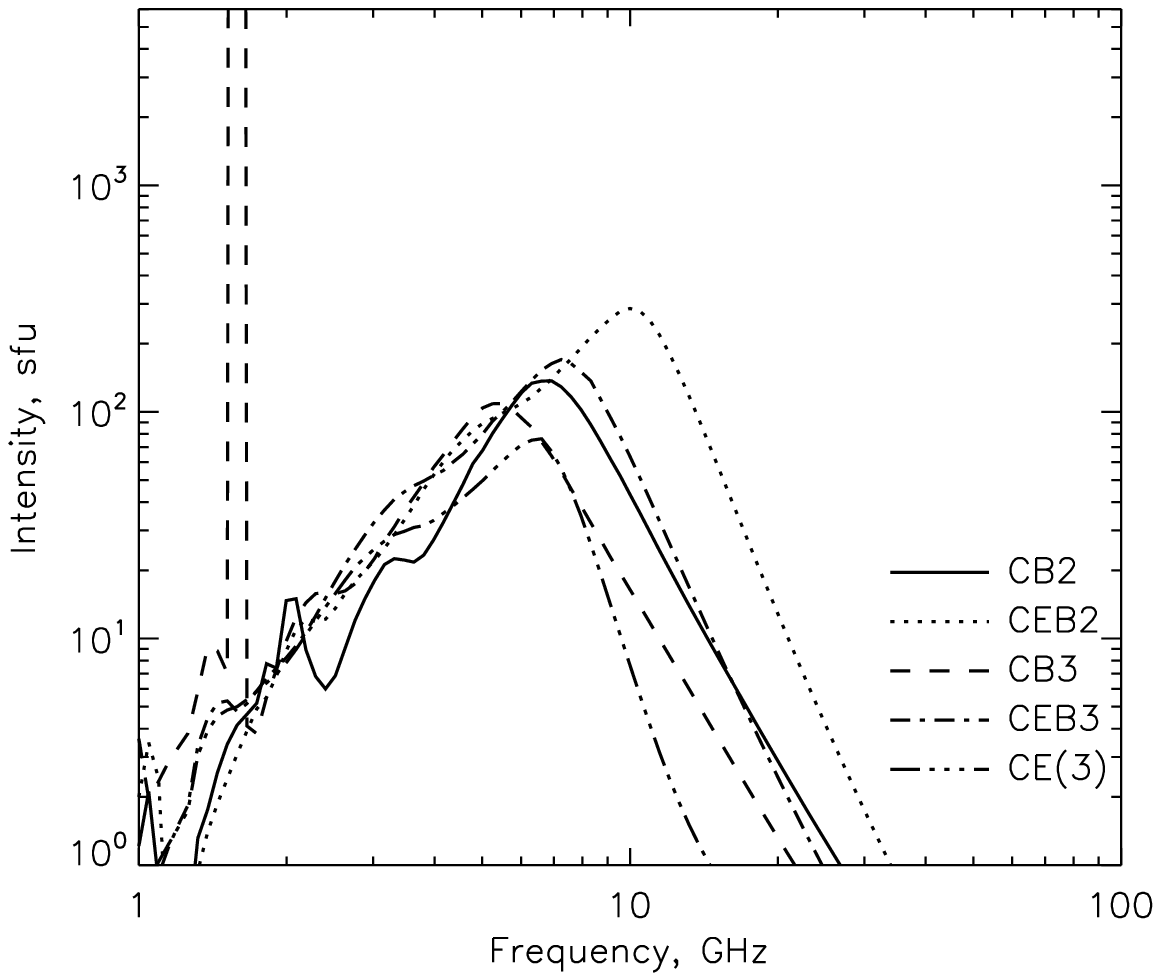}
   \includegraphics[width=6cm]{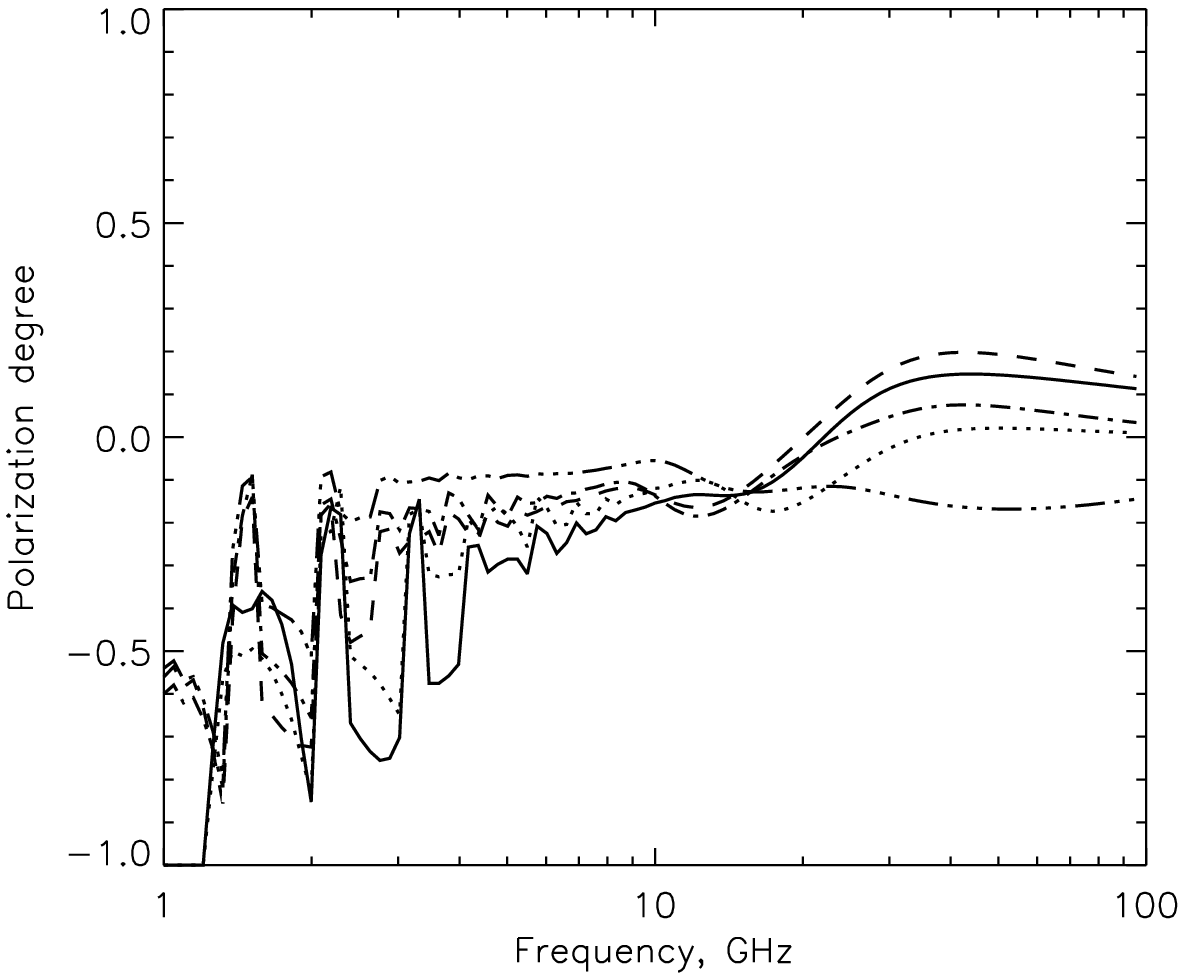}
   \includegraphics[width=6cm]{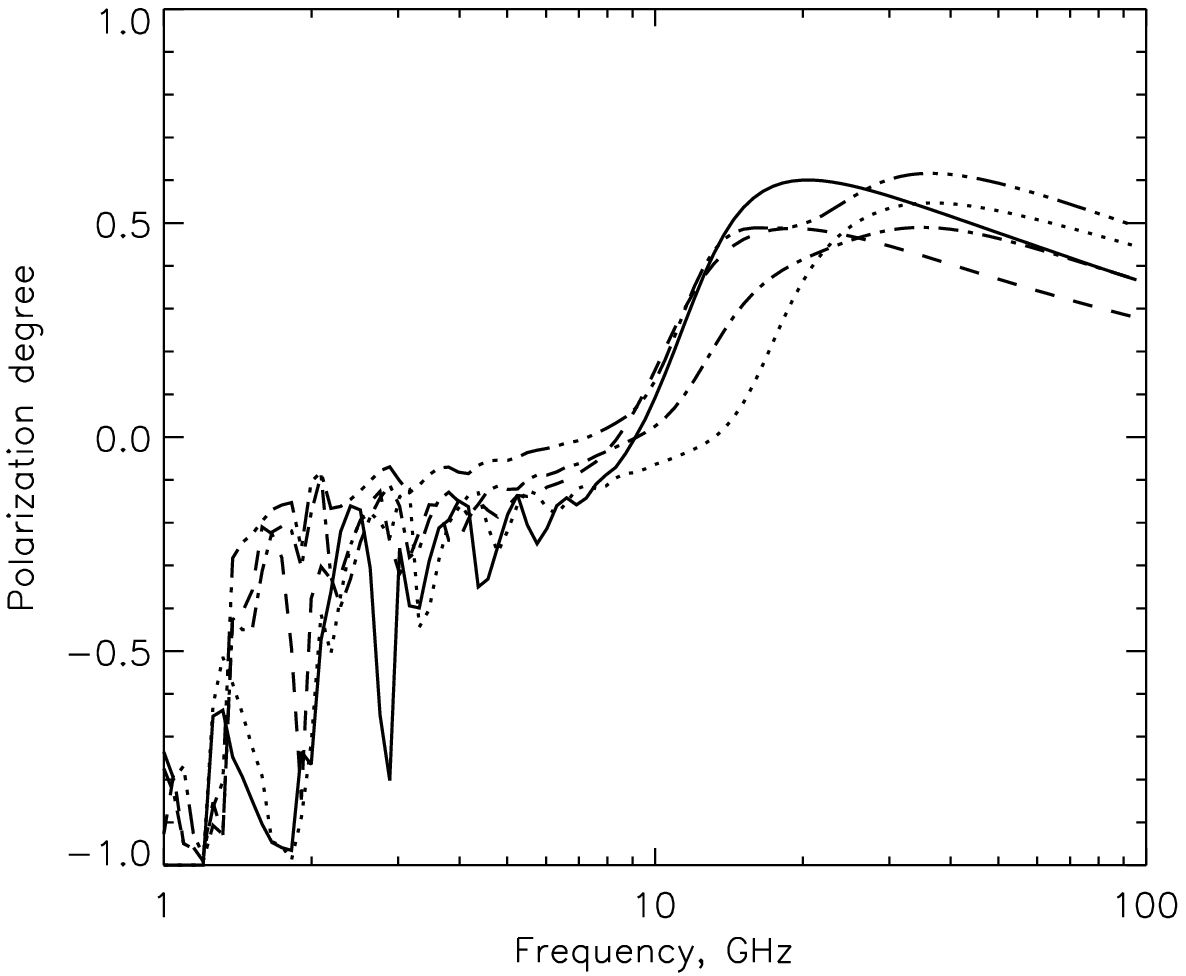}
   \includegraphics[width=6cm]{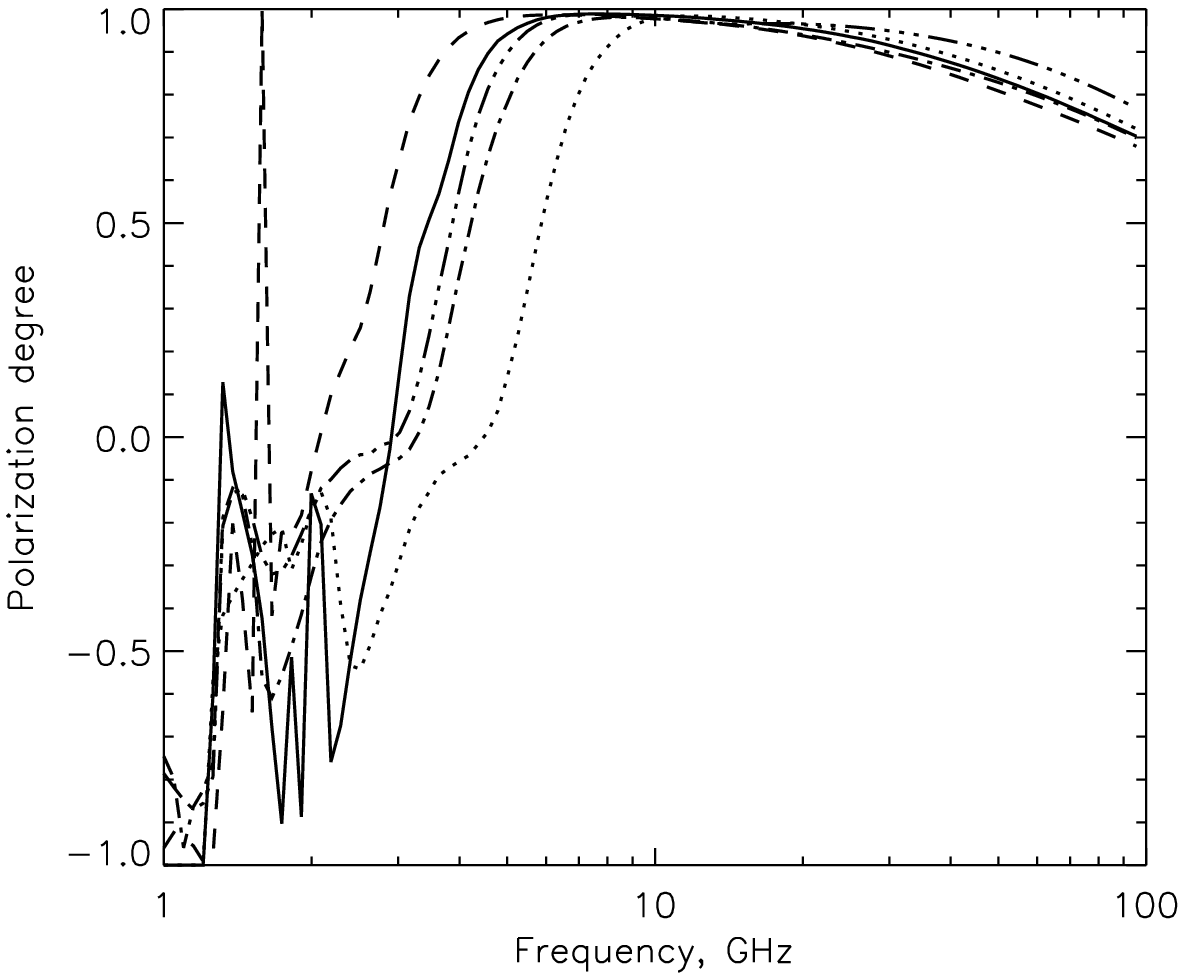}

 \caption{Intensity (equation (\ref{mwint})) in solar flux units (sfu) (upper plots) and degree of polarization (equation (\ref{mwpol})) (lower plots) of the MW emission
spectra calculated for different precipitation models and
propagation directions: for the viewing angles of $110^{\circ}$
(left column plots); $140^{\circ}$ (middle column plots) and
$170^{\circ}$ (right column plots). Numbers in the model
abbreviations define the factor of magnetic field convergence, e.g
CB2 or CEB3 while the number in brackets for CE(3) denotes
simulations for DF calculated for CE model with magnetic field
added in the MW intensity calculations. Note that propagation
direction indicates the viewing angle for the observer looking
from the top and varies from 0 to $90^\circ$ for downward emission
and from $90^\circ$ to $180^\circ$ for upward emission.}
   \label{allMW-models}
 \end{figure*}

MW emission was simulated with radiative transfer approach for
vertically stratified layers for  the following beam precipitation
models described in section~\ref{model} (CB, CE and CEB) and for
different viewing angles (see Fig.~\ref{allMW-models}). It can be
noted that for the viewing angle $\theta=110^{\circ}$, the main
factor affecting MW emission is a magnetic field convergence. MW
intensities have a rather flat maximum at about 20 GHz approaching
the magnitude above $10^3$ sfu for the models with a magnetic
field convergence (CB and CEB models) and reducing the maximum
below $10^3$ sfu at 10 GHz without it (CE model). In the models
including a magnetic field convergence without (CB) or with
electric field (CEB), the plots of MW intensity and polarization
are rather close with the intensities decreasing for a convergence
factor of 3 compared to 2. The models with the converging magnetic
field (CB and CEB) provide much higher emission intensity than the
model without this factor (CE).

The effect of a self-induced electric field is relatively weak and is visible only if the magnetic field convergence is equal to zero, e.g. the model with the return
current (CE) provides a lower peak intensity and a steeper
intensity decrease towards higher frequencies than the collisional model with
magnetic convergence (CB).  The MW intensity at higher energies are power laws with spectral indices of about 2.4 for CEB models and higher than 4 for CE model. The MW frequency distributions towards lower frequencies reveals also power laws with lower spectral index of 2 revealing strong harmonic structure from frequencies of 8 GHz towards zero.

The polarization produced by beams for this viewing angle of $110^{\circ}$ in CE models has always negative sign and ranging in a few percent interval while the polarization produced by CB or CEB models with convergence factor of 2 or 3 is positive (X-mode dominates) for higher frequencies approaching a few percent and crossing zero between 10 and 11 GHz after which it becomes negative (O-mode dominates). The harmonic structure in MW polarization is more pronounced than that in the intensity,  and it increases with a growth of magnetic field convergence.

For a viewing angle $\theta=140^{\circ}$, the effect of a self-induced electric field becomes much more significant than those of the converging magnetic field: the models with collisions and
self-induced electric field (CEB) provide higher MW intensity than the models with collisions and magnetic field
convergence (CB). The maximum MW intensity increases for the models with electric field and magnetic field convergence (CEB) compared to those at $110^\circ$ and frequency of maximum shifts from 11 to 12 GHz for $140^\circ$. The power law spectrum at high frequencies has spectral indices ranging from 2.4 to 2.8 for CEB3 and CEB2 models, respectively. While the lower energy part of MW intensity resembles the power law distributions found for $110^\circ$ with wave-like oscillations with slightly larger amplitudes.

The self-induced electric field has a significant effect on the MW polarization which now changes the sign for all models, and  the frequency of this change shifts to the magnitudes below 10 GHz. The MW polarization degree for CB models is higher than those with electric field (CEB) approaching 50$\%$ at 11 GHz for CB2 model with convergence factor 2 and reducing to 40$\%$ for CB3 with the convergence 3. The polarization for models CEB becomes slightly lower that for CB with rather flat distribution towards higher frequencies and strong wave-like oscillation towards lower frequencies with zero-points between 9 and 10 GHz.

For a viewing angle $170^{\circ}$ the effect of the self-induced electric field is still noticeable for all the models, although the models with collisions, converging magnetic field and
self-induced electric field (CEB) provides slightly higher MW intensity than the models with collisions and electric field (CE) only.However, the maximum MW intensity is significantly (by order of magnitude) reduced for all the models with maxima ranging 7-10 GHz for CE and CEB models, respectively.  The power law spectra at high frequencies become much softer having the spectral indices 4-5. While the lower energy part of MW intensity resembles the power law distributions found for $110^\circ$ with harmonic structure less pronounced than for other viewing angles.

The polarization also changes significantly compared to the other viewing angles approaching about 100$\%$ from frequencies 3-5 GHz for CB3 and CB2 models, shifting to 4-6 for the models CEB3 and CEB2, respectively. The polarization sign change is shifted to the frequencies of 2 GHz for CB3 and 4 GHz for CEB2. The strong harmonic structure with very large amplitudes is still present at lower frequencies.

These distributions of MW intensity and polarization can be understood after considering the combined effect of a self-induced electric field and
converging magnetic field which turns the electrons moving upwards to
such the pitch-angles that their distribution maximum occurs
at about $\mu=-0.8$ \citep{zha06}. Hence, the direction where the
particles emit the MW (and HXR) radiation corresponds to the
intermediate viewing directions ($\theta\simeq
120^{\circ}-150^{\circ}$). As a result, taking into account a
self-induced electric field (in the models CEB) results in the
increase of a maximal MW intensity by a factor 3 (in
comparison with the model CB) and in a noticeable shift of the
spectral peak from 10 to 20-30 GHz.

The spectral index at high frequencies (for $140^\circ$) is -2.4 for the
CEB3 model, and -2.8 for the CEB2 model. This is close to the value
of -3 derived from the observed MW intensity distribution in frequency.
At low frequencies, the spectral index is about 3 that is higher than the
observed one. This can mean that, though the
source is large and inhomogeneous, in the numerical model the largest contribution of the MW emission comes from a compact region near the loop footpoint.

  \begin{figure*}    
 \centerline{\hspace*{0.015\textwidth}
        \includegraphics[width=0.455\textwidth,clip=]{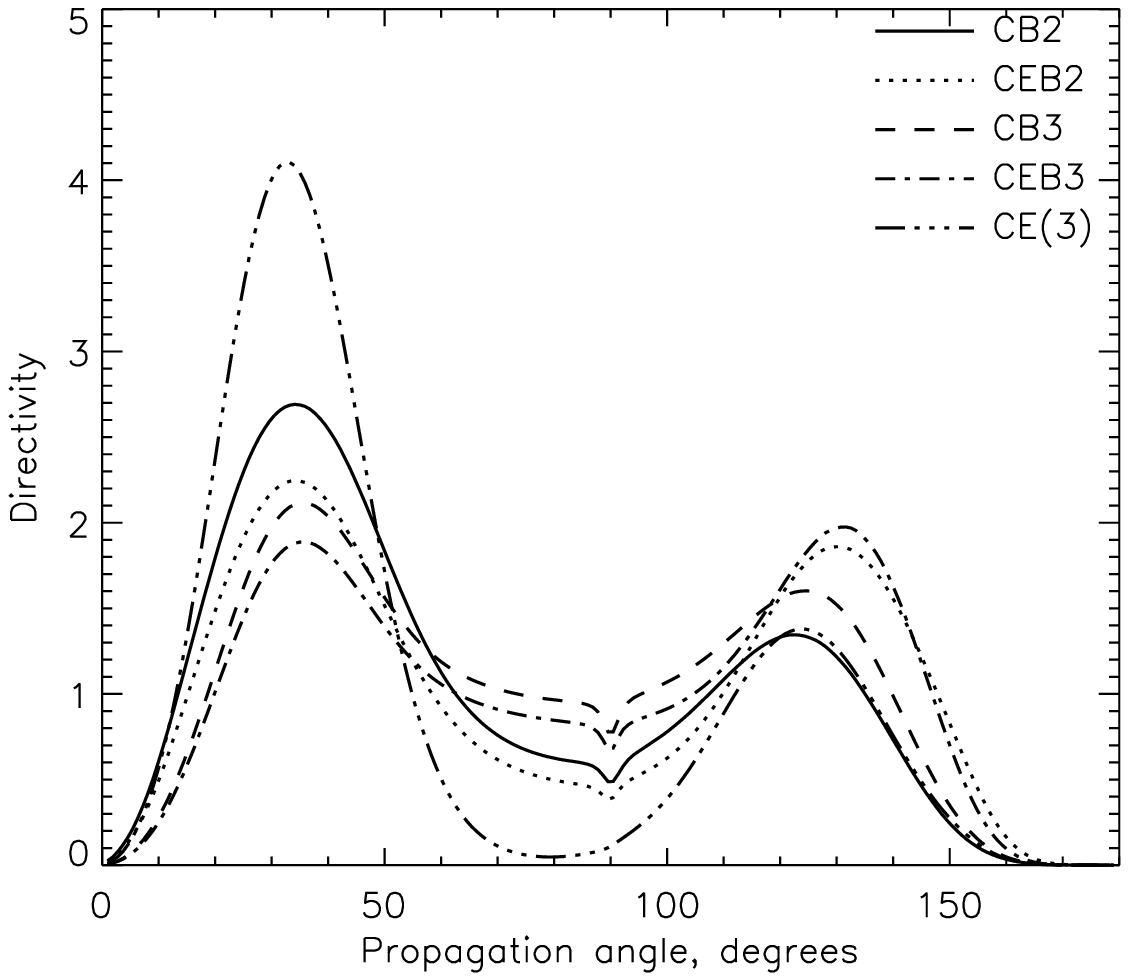}
        \includegraphics[width=0.455\textwidth,clip=]{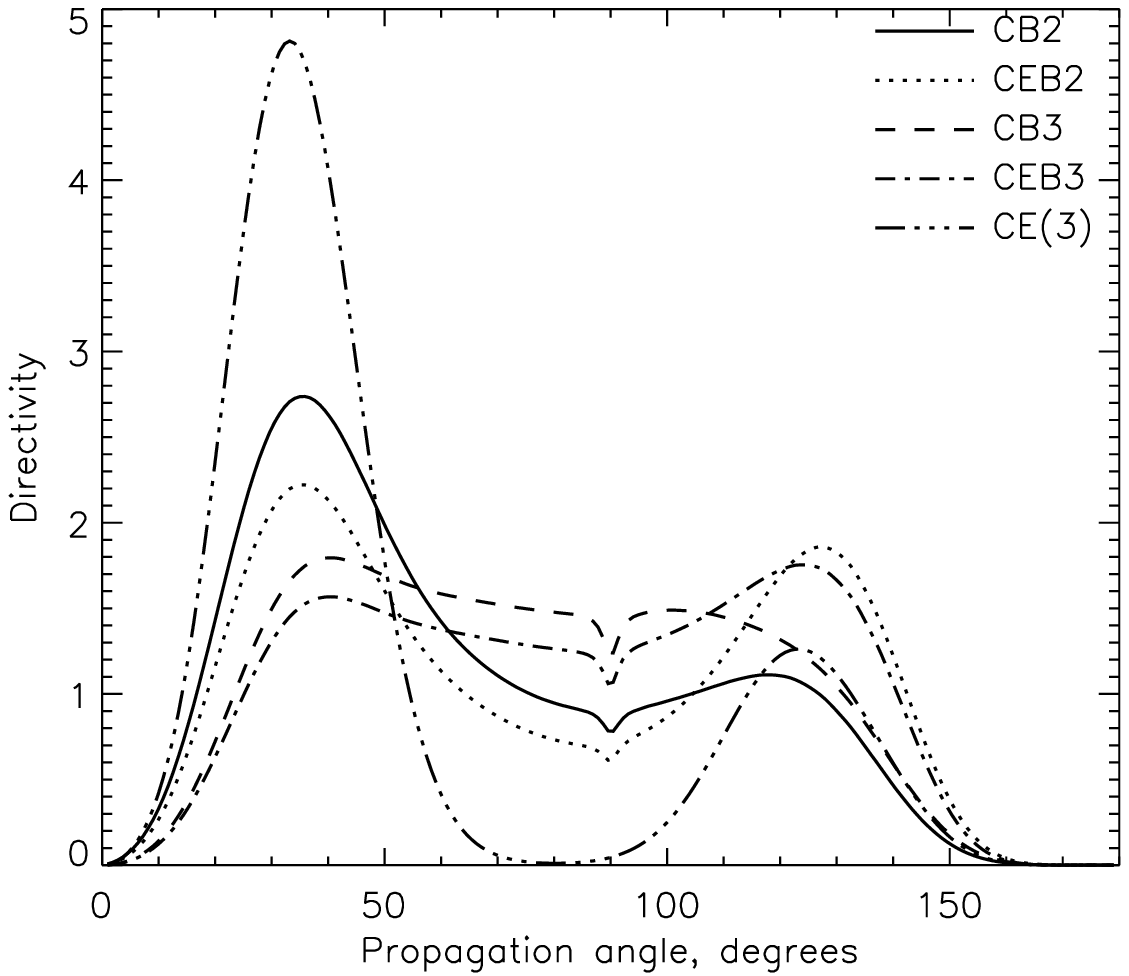}
              }
     \vspace{-0.35\textwidth}   
          \centerline{\large      
         \hfill}
     \vspace{0.33\textwidth}

   \centerline{\hspace*{0.015\textwidth}
               \includegraphics[width=0.455\textwidth,clip=]{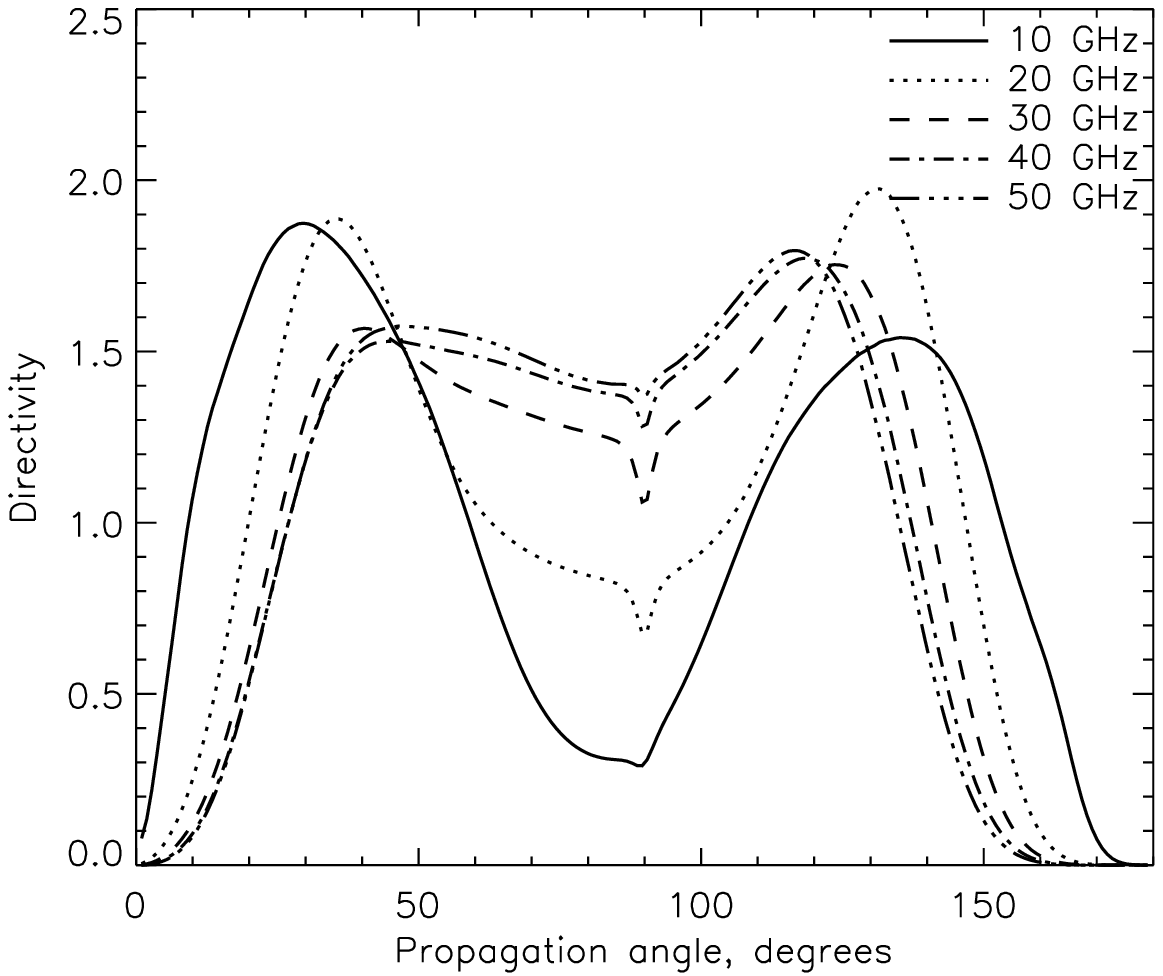}
               \includegraphics[width=0.455\textwidth,clip=]{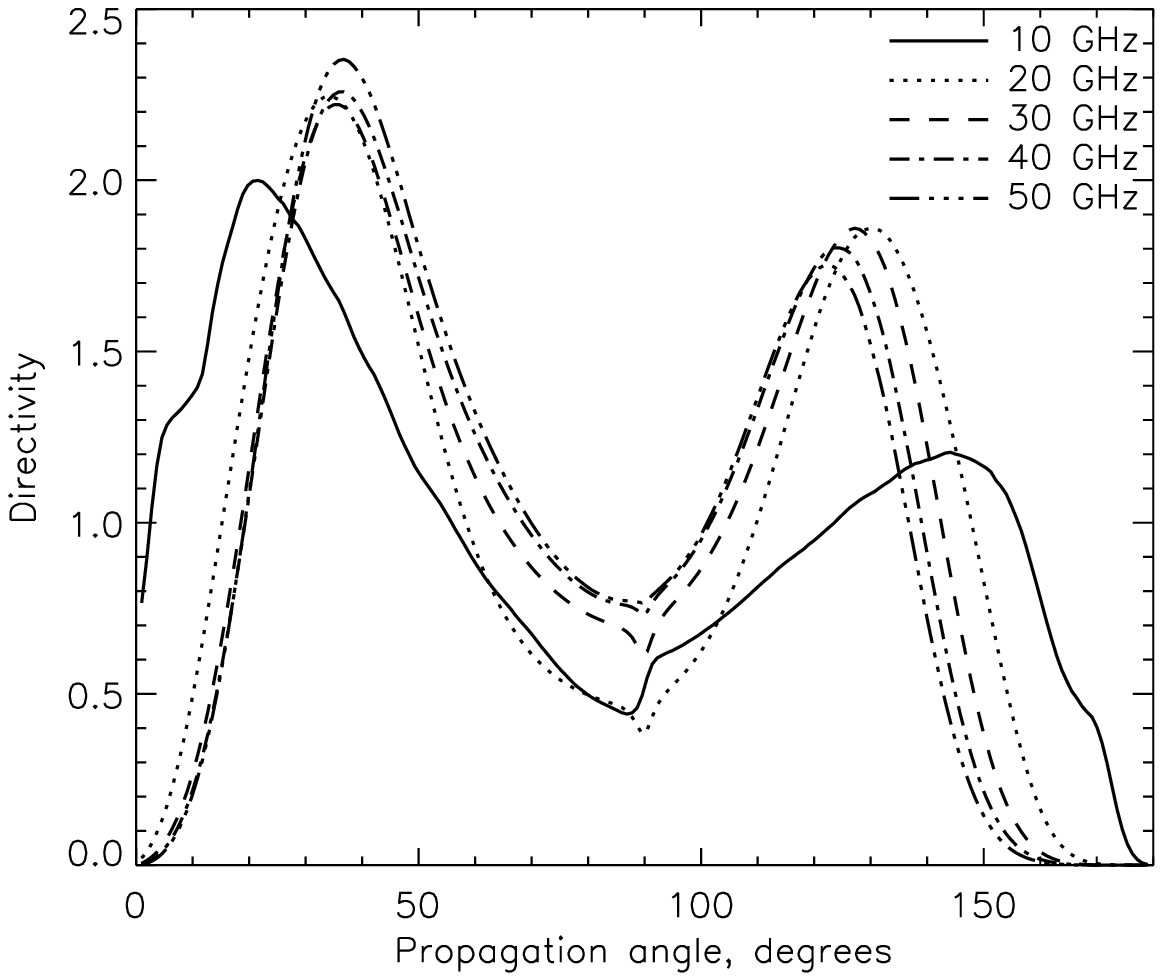}
               \hspace*{0.1mm}
               \hspace*{-0.03\textwidth} {CEB}
            }

\caption{Upper plots: directivity of the integrated MW emission
spectra (equation (\ref{dir}) with the intensity from equation
(\ref{mwint})) versus different propagation directions (see note
in Fig. \ref{allMW-models}) simulated for the models CB, CE, and
CEB (indicated on the plots) with convergence factors 2 and 3 for
20 GHz (left plot) and 30 GHz (right plot). Bottom plots: MW
directivity versus propagation directions calculated for different
frequencies (indicated on the plots)  with CEB3 model (left plot)
and CEB2 model (right plot). CE(3) indicates the MW emission
calculated for the distribution functions with CE model but with
the magnetic field magnitudes at relevant depths corresponding to
CEB3.}
   \label{MW-dir}
\end{figure*}

\subsubsection{MW emission directivity} \label{mw_direct}
Now let us consider the MW directivity plots in
Fig.~\ref{MW-dir}
calculated for the same parameters as MW intensities in
Fig.~\ref{allMW-models}.

It can be noted that the directivity is also strongly dependent on the electron precipitation model applied in calculations. In all models there are two preferential directions for emitting the MW emission: downwards and upwards with the maxima position varying for
different models. For the precipitation models including only electric field (CE(3)) calculated for 20 GHz (left plot) and 30 GHz (right plot) the most of MW emission is emitted in the direction towards the photosphere with a maximum at about $30^\circ$. This maximum is twice higher than the maximum MW emission occurred in the direction of $135^\circ$ (e.g. towards the observer from the top).

For the models including collisions and magnetic field convergence
the directivity of emission for MW emission at 20 GHz  (top left
plot) still has the maxima at the same viewing angle, but the
magnitudes of the maxima are reduced in both directions, although,
keeping the ratio between the downward and upward emission at
about 2 that is close to that for CE model. For the MW emission at
30 GHz (top right plot) the ratio between downward and upward
emission increases to 2.7 for CB2 model approaching 1.4 for CEB2
model. However, CB3 models show some isotropisation of MW emission
between $30^\circ$ and $130^\circ$ where there is no preferential
direction is found while inclusion of the electric field in CEB3
model reveals  again some preferential direction with a small
ratio.

Comparison of the CEB model simulations for different convergence factors (CEB3 in the bottom left plot and CEB2 in the bottom right plot) reveals that directivity is strongest for the model with convergence factor of 2 with the downward-to-upward emission ratio approaching 4, while for the convergence 3 it is only reach 1.5-1.8 for the lowest frequency.
There is also a visible reduction of the directivity with the  increase of frequency for any convergence although, the differences are much higher for the convergence factor of 3 (the left plot). These properties of MW directivity, in addition to intensities and polarization, help us to diagnose electron beam precipitation scenario in this flare.

\subsection{Fit to observations and general discussion} \label{gendis}

The simulations of kinetics of a single electron beam with wide
energy range from 12 keV to 10 MeV  precipitating into a flaring
atmosphere being heated by this beam via hydrodynamic response
is applied to simultaneously interpret the observed MW and HXR emission.

\subsubsection{Hard X-ray emission} \label{hxr_obs}
A comparison of the distributions versus energy of the simulated
and observed HXR photon emission (downward, upwards and their sum)
derived from YOHKOH instrument is presented in Fig.
\ref{HXR-model} for the two viewing angles of $90^\circ$ and
$180^\circ$ and two sets of models: collisions and electric field
(CE) and collisions, electric field and magnetic field convergence
of 3 (CEB). It is obvious that the observed HXR photon spectrum
cannot be fit by any models for the viewing angle of $80^{\circ}$
derived from MW observation by \citet{altyntsev2008}. However, we
can assume that HXR and MW emission come from different parts of
the loop and this discrepancy between the derived viewing angles
can be related to a loop curvature.

\begin{figure*}
 \centering
   \includegraphics[width=8.8cm]{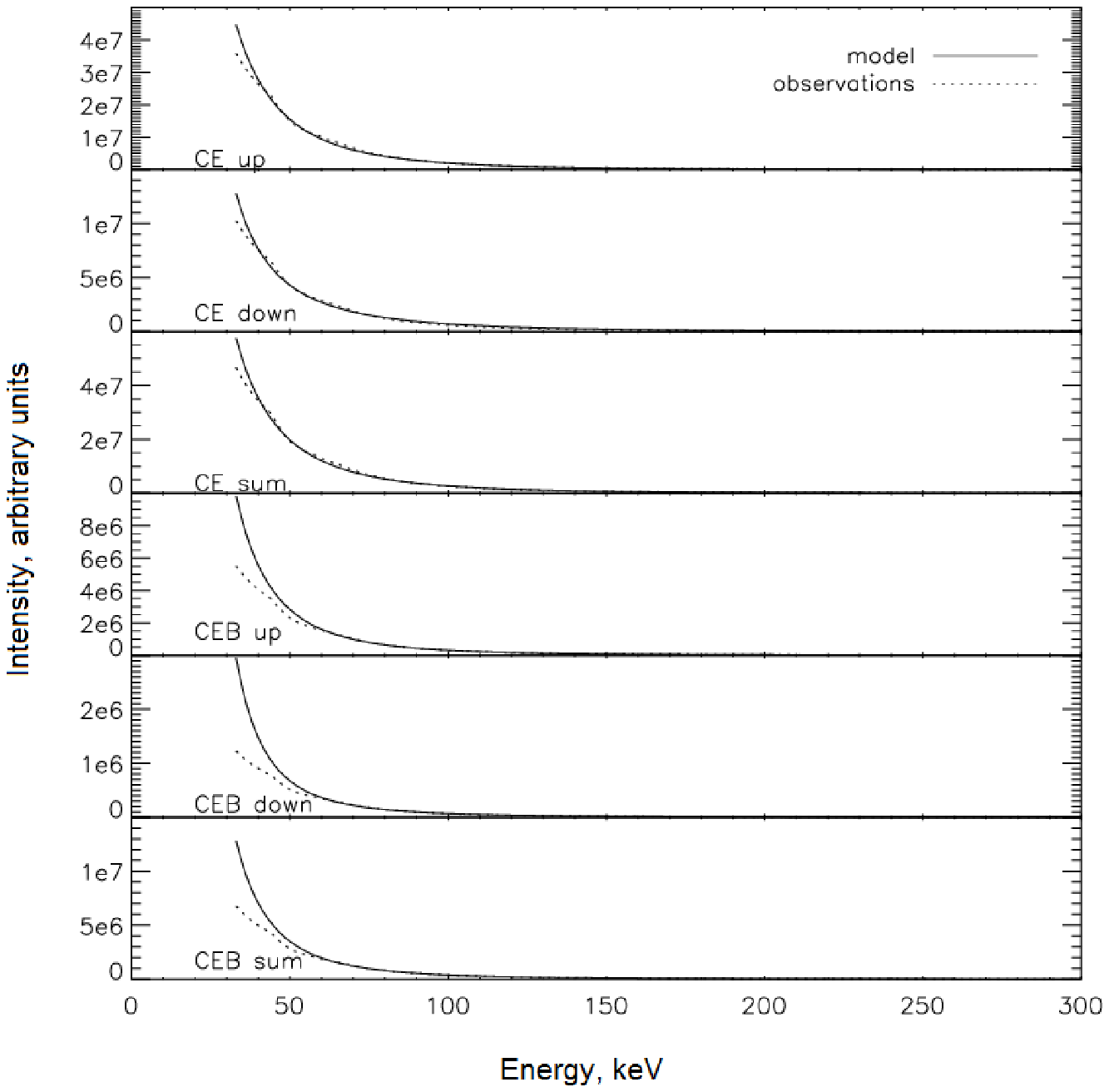}
    \includegraphics[width=9cm]{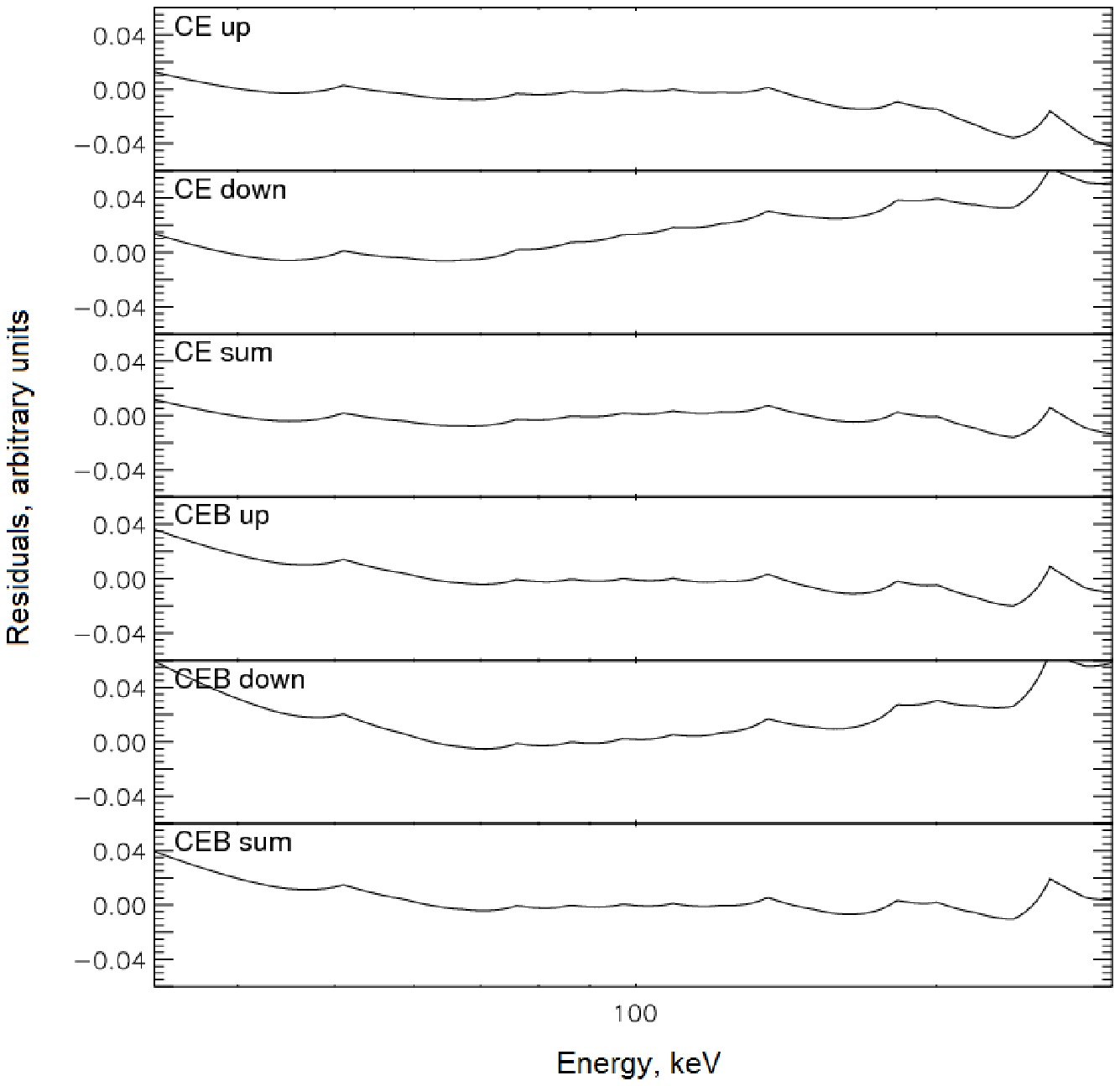}
 \caption{Comparison of the simulated upward, downward and total HXR
 intensities with the  observations (left plots)
 and their residuals (right plots) calculated for
 different beam precipitation models (CE and CEB)
 and the viewing angle $180^\circ$ (along the magnetic field direction).}
   \label{hxr_res}
  \end{figure*}

 Although the agreement between HXR observations by the YOHKOH payload and the proposed models (shown in Figure \ref{hxr_res} in linear scale to amplify the differences) is much better for the simulations carried out for a viewing angle of $180^\circ$ (e.g. upwards along the magnetic field) for the models including electric field (CE and CEB). It is evident from the residuals that the fit is better for CE model (collisions and electric field losses) combining the upward and downward emission (the full albedo effect).

This fit can indicate that, at first, albedo effects are rather important in the interpretation of HXR emission produced by well collimated beams, in addition to isotropic ones proposed earlier \citep{kon06}; and at second, the observed HXR emission is likely to come from the part of a flaring atmosphere not affected by a magnetic field convergence, e.g. from column densities larger then the characteristic column density discussed in section \ref{model}.

\subsubsection{MW emission and polarization} \label{mw_obs}


\begin{table*}
      \caption[]{The $\chi^2$ statistics and significance coefficients for correlation between the observed and simulated MW intensities.}
         \label{coeff}
           \centering
           \begin{tabular}{l c c c c c c c c c}
           \hline
           Points & ${\chi}_0^2$ & ${\chi}_0^2$ &  Kendal  & Spearman  & ${\chi}^2$ & Kendal  & Spearman  & ${\chi}^2$ \\
            of CEB & 0.90\tablefootmark{a}  & 0.99\tablefootmark{b}  & CEB3 \tablefootmark{c}  &   CEB3 \tablefootmark{d}   &    CEB3 \tablefootmark{e}  & CEB2 \tablefootmark{c}   &
CEB2  \tablefootmark{d}    &  CEB2   \tablefootmark{e}  \\
            \hline
            \noalign{\smallskip}
    9 & 3.49 & 1.65 & 0.944 & 0.944 & 3.86&0.889&0.950&2.56    \\
    7 & 2.20 & 0.87 & 0.867 & 0.943 & 2.21&0.733&0.829&0.331   \\
    6 & 1.61 & 0.55 & 1.000 & 1.000 & 1.87&0.800&0.900&0.128   \\
    5 & 1.06  & 0.30 & 1.000 & 1.000 & 1.49&1.000&1.000&0.031    \\
            \noalign{\smallskip}
            \hline

      \end{tabular}
      \tablefoot{
      \tablefoottext{a}{${\chi}_0^2$ for 90$\%$ significantce level from tables by \citet{Chernoff54}; }
      \tablefoottext{b}{${\chi}_0^2$ for 99$\%$ significantce level from tables by \citet{Chernoff54};}
      \tablefoottext{c}{Kendall tau correlation coefficient;}
      \tablefoottext{d}{Spearman correlation coefficient;}
       \tablefoottext{e}{${\chi}^2$ derived from comparison of  the observed and simulated MW intensities for magnetic convergence 2 (CEB2) and 3 (CEB3).}
     }
         \end{table*}

\begin{figure}[h]
\centering
\parbox{0.7\hsize}{
\resizebox{\hsize}{!}{\includegraphics{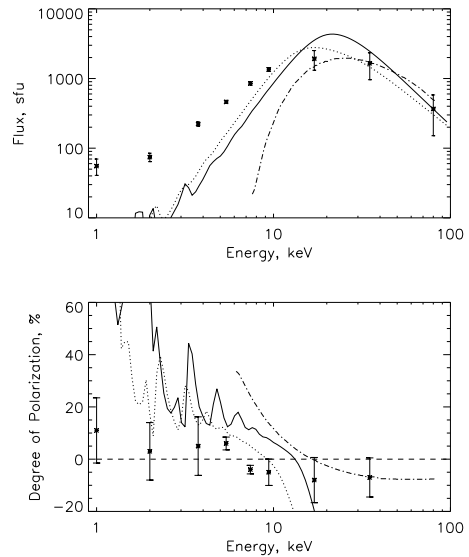}}
}
   \caption{Comparison of observed MW intensity in solar flux units (sfu)(top plot)
    and degree of polarization (bottom plot) with results of the simulations with CEB2
     model for a magnetic field convergence of 2 (dotted line curves) and CEB3 model
      with convergence of 3 (solid line curves) and the model presented by
       \citet{altyntsev2008} (dash-dotted line). In the both plots asterisks mark the
        observed spectra.}
   \label{MW-model}
 \end{figure}

  The distributions in frequency of MW emission  and polarisation
  observed for this flare in  9 frequencies shown by asterisks were  compared
  with the MW emission simulated with a  radiative transfer approach in vertically
stratified layers for different viewing angles (180, 140 and
110$^\circ$) and for different electron precipitation models
including collisions, electric field and magnetic field
convergence as presented in Fig. \ref{MW-model}  for a viewing
angle of 140$^\circ$  revealing the most  reasonable agreement.

The observed MW spectrum  was created by using the NoRP data
plotted in Fig.~\ref{f-norp}a and the MW polarization variations
were taken from the Fig.~\ref{f-norp}b. The MW emission simulated
for optically thin plasma used for fitting the observations in the
paper by \citet{altyntsev2008} is presented on the plot by the
dot-dashed line. The errors of MW emission measurements are shown
by bars in Fig.~\ref{MW-model}. We calculated the standard
deviation of the observed noise level ($\sigma$) for each
frequency of MW observations by SBRS and NoRP. The errors in MW
intensity and polarisation were estimated then from a sum of the
statistical (3$\sigma$) and instrumental errors.

From a hydrodynamic model used to define the temperature and density variations with depth one can estimate the height of a semi-circular loop producing this flare to be
about 10000 km. Then the distance between the MW and HXR intensity
centroids can be estimated at about 10$^{\prime \prime}$ that fits very well the locations of
MW and HXR contours.

The MW emission and polarisation simulated in section \ref{mw_em}
by taking into account radiative transfer effects  reveal smoother
parts at medium and higher frequencies and a harmonic part at
lower frequencies.  The smooth part at higher frequencies has a
negative power law close to that observed while for lower energy
part it has a positive power law with a rather steep slope which
is much higher than the one observed. In the current paper we are
mostly concerned with the smooth part of MW emission and
polarization simulated at higher frequencies for convergence of 2
and 3 (CEB2 and CEB3 and different viewing angles. While at lower
frequencies we have accounted for Razin's effect with the
consideration of depth variations of the ambient density that
improved the MW emission magnitude at lower frequencies. However,
there are still some noticeable disagreements at very low
frequencies between the observed and simulated MW emission as seen
in Fig. \ref{MW-model} . These are possibly, related to the use of
a simplified radiative transfer approach or some other effects
which will require further investigation in the future.

For a viewing angle of 110$^\circ$ , which in our model is equal
to 70$^\circ$ in the model by \citet{altyntsev2008}, the simulated
MW  intensity is not sensitive to the variations of the magnetic
convergence factor and it is smaller than the observed one by more
than 20$\%$ for both lower and higher frequencies. However, for a
viewing angle of $140^\circ$  the model simulations of MW emission
and polarization are able to reproduce very closely the main
features of the observed MW emission. These include the magnitude
and the frequency of MW emission maximum close to the ones
observed; the frequency of the MW spectrum maximum appearing at
about 17 GHz, a negative power law spectral index and a smooth
decrease of MW emission towards lower frequencies with a spectral
index about 2 (being still much higher than the observed one
because of the simplified radiative transfer approach used and
negative power law MW emission with the spectral index of 2.8
towards higher frequencies close to the observed one of 3).

The MW polarization simulated for models CEB2 and CEB3 for the
same viewing angle of $140^\circ$  agrees with the observed
polarisation at lower frequencies  with CEB2 showing a closer
agreement. Also the CEB2 model simulation of MW polarisation
reproduces rather closely (9 GHz), within the limitations of our
radiative transfer model,  the observed frequency (7 GHz) of the
polarisation reversal. While  for higher frequencies the simulated
polarisation is twice higher than the observations that can be the
result of neglecting further scattering of MW emission on the
ambient particles \citep{bast1995,Altyntsev1996}, which can
smoothen significantly a polarisation degree of MW emission at
higher frequencies.

In order to quantify our fits in MW intensities, we carried out a
few statistical tests for the likehood of each curves (CEB2 and
observations, CEB3 and observations) by using the statistical SPSS
package. The tests include Kendall tau and Spearman correlation
tests  \citep{Isobe86, Kendall90} appropriate for other than
normal distributions which we have for MW emission. We also
calculated $\chi^2$ coefficients \citep{Chernoff54} for the
following CEB2 and CEB3 sets: the full number of 9 measurements,
for reduced 7 measurements (excluding the two points at lowest
frequencies), then for 6 and 5 (reducing the measurements
consequently by one from lower frequencies), in order to avoid the
discrepancies in MW emission at the lowest frequencies. The
results of these tests are presented in Table 1. For comparison,
we also include the critical values of $\chi^2$ for given degrees
of freedom corresponding to significance levels of 90$\%$ and
99$\%$.

It can be observed from Table 1 that the intensities for both models (CEB2 and CEB3) reveal strong positive correlation with MW observations for all 9 frequencies, but it increases to a full dependence (correlation coefficients approach 1.0) for the reduced number of measurements of 6 and 5 for CEB3 or 5 for CEB2. Although, for 9 points the correlation is slightly better for the model CEB2. The calculated $\chi^2$ for each datasets (CEB2 and CEB3) from the full to reduced measurement numbers plotted in Table 1 allow us to discriminate between the models. The model CEB3 produces $\chi^2$ of 3.86 for 9 points (or degree of freedom (DF) of 8) that is reduced to 1.49 for 5 points, or DF=4).

The $\chi^2$ coefficients for simulated and observed polarisation
data obtained for  8 observational points (excluding 80 GHz), or 7
degrees of freedom, produced $\chi^2$=1.33 for model CEB2 and 2.13
for model CEB3 that reveals the similar close (above 95$\%$ fit to
observations for the model with convergence 2 because their
$\chi^2$ lie between the critical values for the confidence levels
above 0.90 and below 0.99 \citep{Chernoff54}.  This analysis of
both MW intensity and polarisation  allows to conclude that model
simulations with any convergence fit  reasonably well the MW
observations up to the confidence level of 90$\%$. However, the
model CEB2 with convergence 2 fits the observations up to 95$\%$
for all 9 points and up to the confidence level of 99$\%$ with a
reduced number of 7 points and lower.

Therefore, it is safe to conclude  that with the electron beam
parameters derived from HXR emission, the best agreement with the
MW observations at higher frequencies is achieved for CEB models
with a reasonably small convergence factor of 2, the magnetic
field magnitude at the photosphere derived from MDI (B = 780 G)
and the viewing angle $140^{\circ}$ (see Fig. \ref{MW-model}). The
results of the simulations also allow us to estimate more
accurately the inclination of the loop in the parts emitting the
relevant types of emission: HXR or MW.

One can note that the observed MW emission from this
flare better agrees with our simulated MW emission seen under the viewing angle of $140^\circ$, contrary to the viewing angle of $80^\circ$(equal to 100$^\circ$ in the current model) deduced in the earlier simulations \citep{altyntsev2008} and to the angle  of $180^\circ$ obtained for HXR emission in the current paper. As we discussed above, the difference between the viewing angles  derived for MW emission from our models and those by \citet{altyntsev2008} can be a result of the over-simplified model used in the latter and a limited number of measurements.

However, the difference between MW and HXR viewing angles found in the current paper can be explained within our model by the assumption that the flaring loop was  not  standing perpendicular to the local horizontal plane  in the flare location but was slightly tilted (by about $40^{\circ}$) towards the solar disk center.  By comparing the
directivity of MW and HXR emission from Fig. \ref{dir} one can observe that the MW emission has the preferential direction of 30$^\circ$ downwards and of 130-140$^{\circ}$ upwards.

The downward emission exceeds by factor 2-4 the upward emission. If all these downward photons are fully reflected by the photosphere (albedo effect) with the same properties (angles and energies) they can contribute to reasonable fit of MW emission for the viewing angle of 140$^{\circ}$ compared to 180$^{\circ}$ from HXR spectra, e.g. this difference in viewing angles can be just a result of the directivity effects on HXR and MW intensity from beam electrons. One can also speculate that the loop curvature in the observed flare could contribute to this angle difference for viewing MW and HXR emission to confirm which one needs to extend our model to semi-circular loop that can be done in the  future.

\section{Conclusions} \label{conc}

 In this paper for the first an attempt is made to simulate HXR and MW emission
and polarisation emitted from he flare of 10 March 2001 with the same population of electrons,
in order to improve the fit of observations reported in previous studies. This is achieved by applying the Fokker--Planck kinetic approach to precipitation of electron beam with energy range from 12 keV to 10 MeV into a converging magnetic loop with anisotropic electron scattering on the ambient particles in Coloumb collisions and Ohmic losses.
The theoretical HXR and MW emissions are then calculated by
using these distribution functions for different factors of magnetic field convergence and viewing angles as described in our previous papers \citep{zha10,kuz10}.

The observed HXR photon spectra and frequency distribution of MW
emission and polarisation reveal the best fit for the FP models
including the effect of electric field induced by beam electrons
precipitating in a converging magnetic loop. Magnetic field
strengths in the footpoints on the photosphere were
updated with newly calibrated SOHO/MDI data. The observed HXR
energy spectrum is shown to be a double power law which was fit
very closely by the photon spectrum simulated for the models
including the self-induced electric field.

The MW emission simulated for different models was compared with
the observed distribution in frequency revealing that only the
models combining collisions and electric field effects with pitch
angle anisotropy were able to reproduce closely the main features
of the observed MW emission: peak magnitude and frequency, a
negative spectral index of 2.8 versus 3 observed for higher
frequencies, a smoother decrease of MW emission towards lower
frequencies and the correct frequency of the MW polarisation
reversal. There are still  disagreements between simulated and MW
emission observed at lower frequencies which are likely to be
caused by simplified model of radiative transfer considered in the
current approach and by not taking into account the emission from
the ambient electrons in a vicinity of the flare.

By estimating from the hydrodynamic model the height of a semi-circular loop to be
about 10000 km and by taking into account the inclination of the
loop in HXR emission, the distance between the MW and HXR intensity
centroids was about 10$^{\prime \prime}$ that agrees rather well the
observations.

HXR emission simulated for relativistic angle-dependent cross-section  accounting for different directions of electron
propagation (downwards and upwards) reveal that in the corona the
majority of electrons moves in the downward direction with a twice
smaller number moving upwards while in the chromosphere most
electrons move upwards and only smaller fraction keeps moving
downwards. Thus, for this flare the effects of electron's magnetic mirroring and Ohmic
losses were  significant.

The observed HXR spectrum reveals a noticeable flattening towards lower
energies below 50 keV indicating a significant effect of the self-induced
electric field in beam electron energy losses. The observed HXR photon spectrum was best fit by the model simulated for collisional plus Ohmic losses (CE) precipitation model of electron beam with the initial energy flux of  $10^{12}$ erg/s/cm$^{2}$ and spectral index about 3 including the
emission emitted upwards and downwards (the full albedo effect) in the direction of $180^{\circ}$ (towards the magnetic field direction) that for this flare location indicates a loop tilt about 40$^{\circ}$ towards the solar disk centre.

The observed MW emission distribution in frequency reveals a better  fit for
the model combining collisions and electric field effects with a moderate magnetic field
convergence of 2 (CEB2). This electron precipitation model can reproduce closely within better than $90\%$ confidence level for all 9 points of observations and up to $95\%$ confidence level for 6 points excluding the three ones at lowest frequencies) the main features in the observed MW emission: the maximum magnitude and the frequency at about 17 GHz, a smooth decrease of MW emission towards lower frequencies with a spectral index about 2, the spectral index of 2.8 for higher energy part of MW emission and the frequency of reversal of MW polarisation.

The both models CEB2 and CEB3 show a reasonable fit to the
observed polarisation magnitudes; although the model CEB2 shows a
better fit to the frequency of the MW polarisation reversal close
to the observed one within the $90\%$ confidence level for all 9
points of observations and up to $95\%$ confidence level for 6
points excluding the three observations at lowest frequencies.

In this study the closest fit for observed MW emission was
achieved for the simulations with a weakly (factor 2) converging
magnetic field seen at the viewing  angle of $140^{\circ}$,
contrary to  80$^\circ$ (corresponding to 100$^{\circ}$ in our
model) derived previously \citep{altyntsev2008}. While for HXR
emission the closest agreement between observed and simulated
photon spectra was obtained for the model with collisions and
electric field only with the emission seen at $180^{\circ}$ (from
the loop top).

Within the limitations of our radiative transfer model, this
difference is likely to reflect,  at first, the inclination of
about 40$^\circ$ of the flaring loop towards the solar disk centre
that accounts for the HXR viewing angle. At second, this
difference can also indicate a difference in the preferential
directivity of HXR and  MW emission caused by electron scattering
effects in a presence of self-induced electric and converging
magnetic fields. In addition, a flaring loop  curvature  in the
upper atmosphere where the MW emission is formed can also
contribute to this viewing angle difference.


\begin{acknowledgements}

This study was supported by the Royal Society joint international
project between Bradford University and ISTP, Irkutsk and by the
Russian projects of RFBR No. 09-02-92610,
09-02-00226, 08-02-92204-GFEN, and the RAS Program No. 16. The authors appreciate
the usage of microwave and HXR data obtained with the Yohkoh, the
solar instruments SSRT, NoRP and NoRH. A.A.K. thanks the
Leverhulme Trust for financial support. N.S.M. wishes to thank the
staff of NRO (Japan) for their help and hospitality.
\end{acknowledgements}

\bibliographystyle{aa}
\bibliography{aa_lit}

\begin{thebibliography}{77}
\expandafter\ifx\csname natexlab\endcsname\relax\def\natexlab#1{#1}\fi

\bibitem[{{Altyntsev} {et~al.}(2008){Altyntsev}, {Fleishman}, {Huang}, \&
  {Melnikov}}]{altyntsev2008}
{Altyntsev}, A.~T., {Fleishman}, G.~D., {Huang}, G., \& {Melnikov}, V.~F. 2008,
  \apj, 677, 1367

\bibitem[{{Altyntsev} {et~al.}(1996){Altyntsev}, {Grechnev}, {Konovalov},
  {Lesovoi}, {Lisysian}, {Treskov}, {Rosenraukh}, \& {Magun}}]{Altyntsev1996}
{Altyntsev}, A.~T., {Grechnev}, V.~V., {Konovalov}, S.~K., {et~al.} 1996, \apj,
  469, 976

\bibitem[{{Altyntsev} {et~al.}(2000){Altyntsev}, {Nakajima}, {Takano}, \&
  {Rudenko}}]{alt00}
{Altyntsev}, A.~T., {Nakajima}, H., {Takano}, T., \& {Rudenko}, G.~V. 2000,
  \solphys, 195, 401

\bibitem[{{Aschwanden}(2005)}]{asc05}
{Aschwanden}, M.~J. 2005, {Physics of the Solar Corona. An Introduction with
  Problems and Solutions (2nd edition)}, ed. {Aschwanden, M.~J.}

\bibitem[{{Bastian}(1995)}]{bast1995}
{Bastian}, T.~S. 1995, \apj, 439, 494

\bibitem[{{Bastian}(1999)}]{bas99}
{Bastian}, T.~S. 1999, in Proceedings of the Nobeyama Symposium, held in
  Kiyosato, Japan, Oct. 27-30, 1998, Eds.: T. S. Bastian, N. Gopalswamy and K.
  Shibasaki, NRO Report No. 479., p.211-222, ed. {T.~S.~Bastian, N.~Gopalswamy,
  \& K.~Shibasaki}, 211--222

\bibitem[{{Bastian} {et~al.}(1998){Bastian}, {Benz}, \& {Gary}}]{bas98}
{Bastian}, T.~S., {Benz}, A.~O., \& {Gary}, D.~E. 1998, \araa, 36, 131

\bibitem[{{Battaglia} \& {Benz}(2008)}]{bat2008}
{Battaglia}, M. \& {Benz}, A.~O. 2008, \aap, 487, 337

\bibitem[{{Brown} \& {Bingham}(1984)}]{brown1984aa}
{Brown}, J.~C. \& {Bingham}, R. 1984, \aap, 131, L11

\bibitem[{{Chandra} {et~al.}(2006){Chandra}, {Jain}, {Uddin}, {Yoshimura},
  {Kosugi}, {Sakao}, {Joshi}, \& {Deshpande}}]{chandra2006}
{Chandra}, R., {Jain}, R., {Uddin}, W., {et~al.} 2006, \solphys, 239, 239

\bibitem[{{Chernoff} \& {Lehmann}(1954)}]{Chernoff54}
{Chernoff}, H. \& {Lehmann}, E.~L. 1954, The Annals of Mathematical Statistics,
  25

\bibitem[{{Diakonov} \& {Somov}(1988)}]{dia88}
{Diakonov}, S.~V. \& {Somov}, B.~V. 1988, \solphys, 116, 119

\bibitem[{{Ding}(2003)}]{ding2003}
{Ding}, M.~D. 2003, Journal of Korean Astronomical Society, 36, 49

\bibitem[{{Dulk}(1985)}]{dul85}
{Dulk}, G.~A. 1985, \araa, 23, 169

\bibitem[{{Emslie}(1978)}]{ems78}
{Emslie}, A.~G. 1978, \apj, 224, 241

\bibitem[{{Emslie}(1980)}]{ems80}
{Emslie}, A.~G. 1980, \apj, 235, 1055

\bibitem[{{Fleishman} {et~al.}(2003){Fleishman}, {Gary}, \& {Nita}}]{fle03b}
{Fleishman}, G.~D., {Gary}, D.~E., \& {Nita}, G.~M. 2003, \apj, 593, 571

\bibitem[{{Fleishman} \& {Kuznetsov}(2010)}]{fle2010}
{Fleishman}, G.~D. \& {Kuznetsov}, A.~A. 2010, \apj, 721, 1127

\bibitem[{{Fleishman} \& {Melnikov}(2003)}]{fle03a}
{Fleishman}, G.~D. \& {Melnikov}, V.~F. 2003, \apj, 584, 1071

\bibitem[{{Grechnev} {et~al.}(2008){Grechnev}, {Kurt}, {Chertok}, {Uralov},
  {Nakajima}, {Altyntsev}, {Belov}, {Yushkov}, {Kuznetsov}, {Kashapova},
  {Meshalkina}, \& {Prestage}}]{gre08}
{Grechnev}, V.~V., {Kurt}, V.~G., {Chertok}, I.~M., {et~al.} 2008, \solphys,
  252, 149

\bibitem[{{Hanaoka}(1996)}]{hanaoka96}
{Hanaoka}, Y. 1996, \solphys, 165, 275

\bibitem[{{Hanaoka}(1999{\natexlab{a}})}]{hanaoka99a}
{Hanaoka}, Y. 1999{\natexlab{a}}, \pasj, 51, 483

\bibitem[{{Hanaoka}(1999{\natexlab{b}})}]{hanaoka99b}
{Hanaoka}, Y. 1999{\natexlab{b}}, in Proceedings of the Nobeyama Symposium,
  held in Kiyosato, Japan, Oct. 27-30, 1998, Eds.: T. S. Bastian, N. Gopalswamy
  and K. Shibasaki, NRO Report No. 479., p.229-234, ed. {T.~S.~Bastian,
  N.~Gopalswamy, \& K.~Shibasaki}, 229--234

\bibitem[{{Holman} {et~al.}(2003){Holman}, {Sui}, {Schwartz}, \&
  {Emslie}}]{hol03}
{Holman}, G.~D., {Sui}, L., {Schwartz}, R.~A., \& {Emslie}, A.~G. 2003, \apjl,
  595, L97

\bibitem[{{Isobe} {et~al.}(1986){Isobe}, {Feigelson}, \& {Nelson}}]{Isobe86}
{Isobe}, T., {Feigelson}, E.~D., \& {Nelson}, P.~I. 1986, \apj, 306, 490

\bibitem[{{Kendal} \& {Gobbons}(1990)}]{Kendall90}
{Kendal}, M.~G. \& {Gobbons}, J.~D. 1990, {Rank correlation methods, London:
  Arnold}, 486pp

\bibitem[{{Knight} \& {Sturrock}(1977{\natexlab{a}})}]{kni77}
{Knight}, J.~W. \& {Sturrock}, P.~A. 1977{\natexlab{a}}, \apj, 218, 306

\bibitem[{{Knight} \& {Sturrock}(1977{\natexlab{b}})}]{Knight1977}
{Knight}, J.~W. \& {Sturrock}, P.~A. 1977{\natexlab{b}}, \apj, 218, 306

\bibitem[{{Kontar} {et~al.}(2008){Kontar}, {Hannah}, \& {MacKinnon}}]{kontar08}
{Kontar}, E.~P., {Hannah}, I.~G., \& {MacKinnon}, A.~L. 2008, \aap, 489, L57

\bibitem[{{Kontar} {et~al.}(2006){Kontar}, {MacKinnon}, {Schwartz}, \&
  {Brown}}]{kon06}
{Kontar}, E.~P., {MacKinnon}, A.~L., {Schwartz}, R.~A., \& {Brown}, J.~C. 2006,
  \aap, 446, 1157

\bibitem[{{Kosovichev} \& {Zharkova}(2001)}]{2001ApJ...550L.105K}
{Kosovichev}, A.~G. \& {Zharkova}, V.~V. 2001, \apjl, 550, L105

\bibitem[{{Kosugi} {et~al.}(1991{\natexlab{a}}){Kosugi}, {Makishima}, {Inda},
  {Murakami}, \& {Dotani}}]{kosugi91b}
{Kosugi}, T., {Makishima}, K., {Inda}, M., {Murakami}, T., \& {Dotani}, T.
  1991{\natexlab{a}}, Advances in Space Research, 11, 81

\bibitem[{{Kosugi} {et~al.}(1991{\natexlab{b}}){Kosugi}, {Masuda}, {Makishima},
  {Inda}, {Murakami}, {Dotani}, {Ogawara}, {Sakao}, {Kai}, \&
  {Nakajima}}]{kosugi91a}
{Kosugi}, T., {Masuda}, S., {Makishima}, K., {et~al.} 1991{\natexlab{b}},
  \solphys, 136, 17

\bibitem[{{Krucker} {et~al.}(2008){Krucker}, {Battaglia}, {Cargill},
  {Fletcher}, {Hudson}, {MacKinnon}, {Masuda}, {Sui}, {Tomczak}, {Veronig},
  {Vlahos}, \& {White}}]{kru08}
{Krucker}, S., {Battaglia}, M., {Cargill}, P.~J., {et~al.} 2008, \aapr, 16, 155

\bibitem[{{Krucker} {et~al.}(2010){Krucker}, {Hudson}, {Glesener}, {White},
  {Masuda}, {Wuelser}, \& {Lin}}]{krucker2010}
{Krucker}, S., {Hudson}, H.~S., {Glesener}, L., {et~al.} 2010, \apj, 714, 1108

\bibitem[{{Kundu}(1985)}]{kun85}
{Kundu}, M.~R. 1985, \solphys, 100, 491

\bibitem[{{Kundu} {et~al.}(2001{\natexlab{a}}){Kundu}, {Grechnev}, {Garaimov},
  \& {White}}]{kun01b}
{Kundu}, M.~R., {Grechnev}, V.~V., {Garaimov}, V.~I., \& {White}, S.~M.
  2001{\natexlab{a}}, \apj, 563, 389

\bibitem[{{Kundu} {et~al.}(2009){Kundu}, {Grechnev}, {White}, {Schmahl},
  {Meshalkina}, \& {Kashapova}}]{kun09}
{Kundu}, M.~R., {Grechnev}, V.~V., {White}, S.~M., {et~al.} 2009, \solphys,
  260, 135

\bibitem[{{Kundu} {et~al.}(2004){Kundu}, {Nindos}, \& {Grechnev}}]{kun04}
{Kundu}, M.~R., {Nindos}, A., \& {Grechnev}, V.~V. 2004, \aap, 420, 351

\bibitem[{{Kundu} {et~al.}(2001{\natexlab{b}}){Kundu}, {Nindos}, {White}, \&
  {Grechnev}}]{kun01a}
{Kundu}, M.~R., {Nindos}, A., {White}, S.~M., \& {Grechnev}, V.~V.
  2001{\natexlab{b}}, \apj, 557, 880

\bibitem[{{Kundu} {et~al.}(1995){Kundu}, {Nitta}, {White}, {Shibasaki},
  {Enome}, {Sakao}, {Kosugi}, \& {Sakurai}}]{kun95}
{Kundu}, M.~R., {Nitta}, N., {White}, S.~M., {et~al.} 1995, \apj, 454, 522

\bibitem[{{Kuznetsov} \& {Zharkova}(2010)}]{kuz10}
{Kuznetsov}, A.~A. \& {Zharkova}, V.~V. 2010, \apj, 722, 1577

\bibitem[{{Landau}(1937)}]{lan37}
{Landau}, L.~D. 1937, Zhurn. Experim. Theor.Phys, 7, 203

\bibitem[{{Leach} \& {Petrosian}(1981)}]{lea81}
{Leach}, J. \& {Petrosian}, V. 1981, \apj, 142, 241

\bibitem[{{Lee} \& {Gary}(2000)}]{lee00}
{Lee}, J. \& {Gary}, D.~E. 2000, \apj, 543, 457

\bibitem[{{Lin} {et~al.}(2003){Lin}, {Krucker}, {Hurford}, {Smith}, {Hudson},
  {Holman}, {Schwartz}, {Dennis}, {Share}, {Murphy}, {Emslie}, {Johns-Krull},
  \& {Vilmer}}]{lin03}
{Lin}, R.~P., {Krucker}, S., {Hurford}, G.~J., {et~al.} 2003, \apjl, 595, L69

\bibitem[{{Liu} {et~al.}(2001){Liu}, {Ding}, \& {Fang}}]{Liu2001}
{Liu}, Y., {Ding}, M.~D., \& {Fang}, C. 2001, \apjl, 563, L169

\bibitem[{{Masuda} {et~al.}(1994){Masuda}, {Kosugi}, {Hara}, {Tsuneta}, \&
  {Ogawara}}]{mas94}
{Masuda}, S., {Kosugi}, T., {Hara}, H., {Tsuneta}, S., \& {Ogawara}, Y. 1994,
  \nat, 371, 495

\bibitem[{{McClements}(1992{\natexlab{a}})}]{mcc92a}
{McClements}, K.~G. 1992{\natexlab{a}}, \aap, 253, 261

\bibitem[{{McClements}(1992{\natexlab{b}})}]{mcc92b}
{McClements}, K.~G. 1992{\natexlab{b}}, \aap, 258, 542

\bibitem[{{Melnikov} {et~al.}(2008){Melnikov}, {Gary}, \& {Nita}}]{Melnikov08}
{Melnikov}, V.~F., {Gary}, D.~E., \& {Nita}, G.~M. 2008, \solphys, 253, 43

\bibitem[{{Melrose}(1968)}]{melrose68}
{Melrose}, D.~B. 1968, \apss, 2, 171

\bibitem[{{Melrose}(1999)}]{mel99}
{Melrose}, D.~B. 1999, in Proceedings of the Nobeyama Symposium, held in
  Kiyosato, Japan, Oct. 27-30, 1998, Eds.: T. S. Bastian, N. Gopalswamy and K.
  Shibasaki, NRO Report No. 479., p.371-380, ed. {T.~S.~Bastian, N.~Gopalswamy,
  \& K.~Shibasaki}, 371--380

\bibitem[{{Nakajima} {et~al.}(1994){Nakajima}, {Nishio}, {Enome}, {Shibasaki},
  {Takano}, {Hanaoka}, {Torii}, {Sekiguchi}, {Bushimata}, {Kawashima},
  {Shinohara}, {Irimajiri}, {Koshiishi}, {Kosugi}, {Shiomi}, {Sawa}, \&
  {Kai}}]{nakajima94}
{Nakajima}, H., {Nishio}, M., {Enome}, S., {et~al.} 1994, IEEE Proceedings, 82,
  705

\bibitem[{{Nakajima} {et~al.}(1985){Nakajima}, {Sekiguchi}, {Sawa}, {Kai}, \&
  {Kawashima}}]{nakajima}
{Nakajima}, H., {Sekiguchi}, H., {Sawa}, M., {Kai}, K., \& {Kawashima}, S.
  1985, \pasj, 37, 163

\bibitem[{{Nita} {et~al.}(2004){Nita}, {Gary}, \& {Lee}}]{Nita04}
{Nita}, G.~M., {Gary}, D.~E., \& {Lee}, J. 2004, \apj, 605, 528

\bibitem[{{Ramaty}(1969)}]{ramaty69}
{Ramaty}, R. 1969, \apj, 158, 753

\bibitem[{{Sato} {et~al.}(2006){Sato}, {Matsumoto}, {Yoshimura}, {Kubo},
  {Kotoku}, {Masuda}, {Sawa}, {Suga}, {Yoshimori}, {Kosugi}, \&
  {Watanabe}}]{sato}
{Sato}, J., {Matsumoto}, Y., {Yoshimura}, K., {et~al.} 2006, \solphys, 236, 351

\bibitem[{{Scherrer} {et~al.}(1995){Scherrer}, {Bogart}, {Bush}, {Hoeksema},
  {Kosovichev}, {Schou}, {Rosenberg}, {Springer}, {Tarbell}, {Title},
  {Wolfson}, {Zayer}, \& {MDI Engineering Team}}]{scherrer}
{Scherrer}, P.~H., {Bogart}, R.~S., {Bush}, R.~I., {et~al.} 1995, \solphys,
  162, 129

\bibitem[{{Shibasaki} {et~al.}(1979){Shibasaki}, {Ishiguro}, \&
  {Enome}}]{shibasaki}
{Shibasaki}, K., {Ishiguro}, M., \& {Enome}, S. 1979, Nagoya University,
  Research Institute of Atmospherics, Proceedings, vol.~26, Mar.~1979,
  p.~117-127., 26, 117

\bibitem[{{Siversky} \& {Zharkova}(2009)}]{siversky09}
{Siversky}, T.~V. \& {Zharkova}, V.~V. 2009, \aap, 504, 1057

\bibitem[{{Sudol} \& {Harvey}(2005)}]{2005ApJ...635..647S}
{Sudol}, J.~J. \& {Harvey}, J.~W. 2005, \apj, 635, 647

\bibitem[{{Sui} {et~al.}(2002){Sui}, {Holman}, {Dennis}, {Krucker}, {Schwartz},
  \& {Tolbert}}]{sui02}
{Sui}, L., {Holman}, G.~D., {Dennis}, B.~R., {et~al.} 2002, \solphys, 210, 245

\bibitem[{{Syrovatskii} \& {Shmeleva}(1972)}]{Syrovatsky72}
{Syrovatskii}, S.~I. \& {Shmeleva}, O.~P. 1972, \sovast, 16, 273

\bibitem[{{Takakura} {et~al.}(1995){Takakura}, {Kosugi}, {Sakao}, {Makishima},
  {Inda-Koide}, \& {Masuda}}]{tak95}
{Takakura}, T., {Kosugi}, T., {Sakao}, T., {et~al.} 1995, \pasj, 47, 355

\bibitem[{{Torii} {et~al.}(1979){Torii}, {Tsukiji}, {Kobayashi}, {Yoshimi},
  {Tanaka}, \& {Enome}}]{torii}
{Torii}, C., {Tsukiji}, Y., {Kobayashi}, S., {et~al.} 1979, Nagoya University,
  Research Institute of Atmospherics, Proceedings, vol.~26, Mar.~1979,
  p.~129-132., 26, 129

\bibitem[{{Uddin} {et~al.}(2004){Uddin}, {Jain}, {Yoshimura}, {Chandra},
  {Sakao}, {Kosugi}, {Joshi}, \& {Despande}}]{uddin2004}
{Uddin}, W., {Jain}, R., {Yoshimura}, K., {et~al.} 2004, \solphys, 225, 325

\bibitem[{{van den Oord}(1990)}]{oord90}
{van den Oord}, G.~H.~J. 1990, \aap, 234, 496

\bibitem[{{Vilmer} {et~al.}(2002){Vilmer}, {Krucker}, {Lin}, \& {The Rhessi
  Team}}]{vil02}
{Vilmer}, N., {Krucker}, S., {Lin}, R.~P., \& {The Rhessi Team}. 2002,
  \solphys, 210, 261

\bibitem[{{Wilson} \& {Holman}(2003)}]{wil03}
{Wilson}, R.~F. \& {Holman}, G.~D. 2003, AAS SPD meeting, 34, 1620

\bibitem[{{Yoshimori} {et~al.}(1991){Yoshimori}, {Okudaira}, {Hirasima},
  {Igarashi}, {Akasaka}, {Takai}, {Morimoto}, {Watanabe}, {Ohki}, \&
  {Nishimura}}]{yoshimori}
{Yoshimori}, M., {Okudaira}, K., {Hirasima}, Y., {et~al.} 1991, \solphys, 136,
  69

\bibitem[{{Zharkova} {et~al.}(1995){Zharkova}, {Brown}, \&
  {Syniavskii}}]{zha95}
{Zharkova}, V.~V., {Brown}, J.~C., \& {Syniavskii}, D.~V. 1995, \aap, 304, 284

\bibitem[{{Zharkova} \& {Gordovskyy}(2006)}]{zha06}
{Zharkova}, V.~V. \& {Gordovskyy}, M. 2006, \apj, 651, 553

\bibitem[{{Zharkova} {et~al.}(2011){Zharkova}, {Kashapova}, {Chornogor}, \&
  {Andrienko}}]{Zharkova11}
{Zharkova}, V.~V., {Kashapova}, L.~K., {Chornogor}, S., \& {Andrienko}, O.
  2011, Monthly Notices of the Royal Astronomical Society, in press

\bibitem[{{Zharkova} {et~al.}(2010){Zharkova}, {Kuznetsov}, \&
  {Siversky}}]{zha10}
{Zharkova}, V.~V., {Kuznetsov}, A.~A., \& {Siversky}, T.~V. 2010, \aap, 512,
  A8+

\bibitem[{{Zharkova} \& {Zharkov}(2007)}]{Zharkova07}
{Zharkova}, V.~V. \& {Zharkov}, S.~I. 2007, \apj, 664, 573

\bibitem[{{Zharkova} {et~al.}(2005){Zharkova}, {Zharkov}, {Ipson}, \&
  {Benkhalil}}]{Zharkova2005JGR}
{Zharkova}, V.~V., {Zharkov}, S.~I., {Ipson}, S.~S., \& {Benkhalil}, A.~K.
  2005, Journal of Geophysical Research (Space Physics), 110, A08104

\end{thebibliography}
\end{document}